\documentstyle[aps,preprint,tighten,epsf,eqsecnum,floats]{revtex}

\begin{document}
\title{Diagrammatic method of integration over the unitary group, \\ with applications to quantum transport in mesoscopic systems}
\author{P. W. Brouwer and C. W. J. Beenakker}
\address{Instituut-Lorentz, University of Leiden, P.O. Box 9506, 2300 RA
Leiden, The Netherlands}
\maketitle

\begin{abstract}
A diagrammatic method is presented for averaging over the circular ensemble of random-matrix theory. The method is applied to phase-coherent conduction through a chaotic cavity (a ``quantum dot'') and through the interface between a normal metal and a superconductor. \\
\bigskip
\end{abstract}
\newpage

\section{Introduction}

The random-matrix theory of quantum transport describes the statistics of transport properties of phase-coherent (mesoscopic) systems in terms of the statistics of random matrices (for reviews, see Refs.\ \onlinecite{WeidenmuellerReview,StoneReview,MelloReview,BeenakkerReview}). There exist two separate (but equivalent) approaches: Either the random matrix is used to model the Hamiltonian of the closed system, or it is used to model the scattering matrix of the open system. The second approach is more direct than the first, because the scattering matrix directly determines the conductance through the Landauer formula,
\begin{equation} \label{eq:Landauer}
  G = {2e^2 \over h} \mbox{tr}\, t t^{\dagger}.
\end{equation}
(The transmission matrix $t$ is a submatrix of the scattering matrix.)

Random-matrix theory has been applied successfully to two types of mesoscopic systems: chaotic cavities and disordered wires. Baranger and Mello \cite{BarangerMello} and Jalabert, Pichard, and Beenakker \cite{JPB} studied conduction through a chaotic cavity on the assumption that the scattering matrix $S$ is uniformly distributed in the unitary group, restricted only by symmetry. This is the circular ensemble, introduced by Dyson \cite{Dyson}, and shown to apply to a chaotic cavity by Bl\"umel and Smilansky \cite{BlumelSmilansky}. The symmetry restriction is that $S S^{*} = 1$ in the presence of time-reversal symmetry. (The superscript $*$ indicates complex conjugation if the elements of $S$ are complex numbers; in the presence of spin-orbit scattering, $S$ is a matrix of quaternions, and $S^{*}$ denotes the quaternion complex conjugate.) For the disordered wire, the circular ensemble applies not to the scattering matrix itself, but to the unitary matrices $v$, $w$, $v'$, and $w'$ in the polar decomposition,
\begin{equation}
  \label{eq:polardecomp1}
  S = \left( \begin{array}{cc} v & 0 \\ 0 & w \end{array} \right)
      \left( \begin{array}{cc} \sqrt{1-T} & i \sqrt{T} \\ 
        i \sqrt{T} & \sqrt{1-T} \end{array} \right)
      \left( \begin{array}{cc} v' & 0 \\ 
        0 & w' \end{array} \right).
\end{equation}
The matrix $T$ is a diagonal matrix containing the transmission eigenvalues $T_{n} \in [0,1]$ on the diagonal. (The $T_{n}$'s are the eigenvalues of the matrix product $t t^{\dagger}$.) The distribution of the transmission eigenvalues is governed by a Fokker-Planck equation, the Dorokhov-Mello-Pereyra-Kumar (DMPK) equation \cite{Dorokhov,MPK}. The isotropy assumption \cite{MPK} states that $v$, $v'$, $w$, and $w'$ are uniformly and independently distributed in the unitary group, with the restriction $v^{*} v' = 1$, $w^{*} w' = 1$ in the presence of time-reversal symmetry.

The role of the circular ensemble of unitary matrices in the scattering matrix approach is comparable to the role of the Gaussian ensemble of Hermitian matrices in the Hamiltonian approach. However, whereas many computational techniques have been developed for averaging over the Gaussian ensemble \cite{Pastur,Pandey,Efetov,VWZ,IWZ,PrigodinEfetovIida,BrezinZee,MMZ}, the circular ensemble has received less attention. If the dimension $N$ of the unitary matrices is small, the average over the circular ensemble can be done exactly \cite{BrouwerBeenakker94,BarangerMello96}. For some applications in the regime of large $N$, one may regard the elements of the unitary matrix as independent Gaussian variables \cite{FriedmanMello2}, and then use the known diagrammatic perturbation theory for the Gaussian ensemble \cite{Pandey,BrezinZee}. In other applications the Gaussian approximation breaks down.

In this paper we present a diagrammatic technique for integration over the unitary group, which is not restricted to the Gaussian approximation. We discuss two applications: A chaotic cavity coupled to the outside via a tunnel barrier, and a disordered wire attached to a superconductor. In both cases, we calculate the mean and variance of the conductance up to and including terms of order $1$. We point out the analogy between the diagrams contributing to the average over the circular ensemble and the diffuson and cooperon diagrams which appear in the theory of weak localization \cite{Anderson,Gorkov} and universal conductance fluctuations \cite{Altshuler,LeeStone} in disordered metals. In the presence of the superconductor a third type of diagrams shows up, which gives rise to the coexistence of weak localization with a magnetic field \cite{BrouwerBeenakker95-1,AltlandZirnbauer}, and to anomalous conductance fluctuations \cite{BrouwerBeenakker95-2}. 

The paper starts in Sec.\ \ref{sec:UAVG} with a summary of known results \cite{Creutz,Samuel,Mello} for the integration over the unitary group of a polynomial function of matrix elements. The diagrammatic technique is explained in Sec.\ \ref{sec:DIAGR}. Generalizations to unitary symmetric matrices and to unitary quaternion matrices are given in Secs.\ \ref{sec:UAVGb1} and \ref{sec:trans}, respectively. We then apply the technique to the chaotic cavity (Sec.\ \ref{sec:QD}) and the normal-metal--superconductor junction (Sec.\ \ref{sec:NS}). Some of the results of Sec.\ \ref{sec:QD} have been obtained previously by Iida, Weidenm\"uller, and Zuk, who used the Hamiltonian approach to quantum transport and the supersymmetry technique \cite{WeidenmuellerReview,IWZ}. The results of Sec.\ \ref{sec:NS} have been published in Refs.\ \onlinecite{BrouwerBeenakker95-1,BrouwerBeenakker95-2}, without the detailed derivation presented here. There is some overlap between Sec.\ \ref{sec:NS} and a recent work by Argaman and Zee \cite{Argaman}.

\section{Integration of polynomials of unitary matrices} \label{sec:UAVG}

In this section we summarize known results \cite{Creutz,Samuel,Mello} for the integration of a polynomial function $f(U)$ of the matrix elements of an $N \times N$ unitary matrix $U$ over the unitary group ${\cal U}(N)$. We refer to the integration as an ``average'', which we denote by brackets $\langle \cdots \rangle$, 
\begin{equation}
  \langle f \rangle \equiv \int dU\, f(U).
\end{equation}
Here $dU$ is the invariant measure (Haar measure) on ${\cal U}(N)$, normalized to unity ($\int dU = 1$). The ensemble of unitary matrices that corresponds to this average is known as the circular unitary ensemble (CUE) \cite{Dyson,Mehta}.

We consider a polynomial function $f(U) = U^{\vphantom{*}}_{a_1 b_1} \ldots U^{\vphantom{*}}_{a_n b_n} U^{*}_{\alpha_1 \beta_1} \ldots U^{*}_{\alpha_m \beta_m}$. The average $\langle f(U) \rangle$ is zero unless $n=m$, $\alpha_1,\ldots,\alpha_n$ is a permutation $P$ of $a_1,\ldots,a_n$, and $\beta_1,\ldots,\beta_n$ is a permutation $P'$ of $b_1,\ldots,b_n$. The general structure of the average is
\begin{eqnarray} \label{eq:Uperm} \label{eq:Uavg}
 \left\langle U^{\vphantom{*}}_{a_1 b_1} \ldots U^{\vphantom{*}}_{a_m b_m} 
    U^{*}_{\alpha_1 \beta_1} \ldots U^{*}_{\alpha_n \beta_n} \right\rangle = 
  \delta_{n m}
  \sum_{P,P'} V_{P,P'} \prod_{j=1}^{n}
    \delta_{a_j \alpha_{P(j)}} \delta_{b_j \beta_{P'(j)}},
\end{eqnarray}
where the summation is over all permutations $P$ and $P'$ of the numbers $1, \ldots, n$. The coefficients $V_{P,P'}$ depend only on the {\em cycle structure} of the permutation $P^{-1} P'$ \cite{Samuel}. Recall that each permutation of $1,\ldots,n$ has a unique factorization in disjoint cyclic permutations (``cycles'') of lengths $c_1, \ldots, c_k$ (where $n = \sum_{j=1}^{k} c_k$). The statement that $V_{P,P'}$ depends only on the cycle structure of $P^{-1} P'$ means that $V_{P,P'}$ depends only on the lengths $c_1, \ldots, c_k$ of the cycles in the factorization of $P^{-1} P'$. One may therefore write $V_{c_1, \ldots, c_k}$ instead of $V_{P,P'}$.

As an example, we consider the case $n=m=2$ explicitly. The summation over the permutations $P$ and $P'$ extends over the identity permutation $\mbox{\it id} = [(1,2) \to (1,2)]$ and the exchange permutation $\mbox{\it ex} =[(1,2) \to (2,1)]$. Hence Eq.\ (\ref{eq:Uperm}) reads
\begin{eqnarray} \label{eq:UpermEx1}
  \left\langle U^{\vphantom{*}}_{a_1 b_1} U^{\vphantom{*}}_{a_2 b_2} 
    U^{*}_{\alpha_1 \beta_1} U^{*}_{\alpha_2 \beta_2} \right\rangle &=& 
  \lefteqn{V_{{id},{id}}}\hphantom{V_{{id},{ex}}}\,
          \delta_{a_1 \alpha_1} \delta_{b_1 \beta_1}
          \delta_{a_2 \alpha_2} \delta_{b_2 \beta_2} +
  \lefteqn{V_{{ex},{id}}}\hphantom{V_{{ex},{ex}}}\,
          \delta_{a_1 \alpha_2} \delta_{b_1 \beta_1}
          \delta_{a_2 \alpha_1} \delta_{b_2 \beta_2} \nonumber \\ && \mbox{} +
  V_{{id},{ex}}\, \delta_{a_1 \alpha_1} \delta_{b_1 \beta_2}
          \delta_{a_2 \alpha_2} \delta_{b_2 \beta_1} +
  V_{{ex},{ex}}\, \delta_{a_1 \alpha_2} \delta_{b_1 \beta_2}
          \delta_{a_2 \alpha_1} \delta_{b_2 \beta_1}.
\end{eqnarray}
The permutation $P^{-1} P'$ that corresponds to $P=P'=\mbox{\it id}$ [the first term on the r.h.s.\ of Eq.\ (\ref{eq:UpermEx1})] is again the identity permutation: $P^{-1} P' = \mbox{\it id} = [(1,2) \to (1,2)]$. Its factorization in cyclic permutations is $\mbox{\it id} = (1 \to 1)(2 \to 2)$, so that $P^{-1} P'$ factorizes in two cyclic permutations of unit length. Hence the cycle structure of $P^{-1} P'$ is $\{1,1\}$, and $V_{id,id} = V_{1,1}$. The second term on the r.h.s.\ of Eq.\ (\ref{eq:UpermEx1}), corresponding to $P=\mbox{\it ex}$, $P'=\mbox{\it id}$, has $P^{-1} P' = \mbox{\it ex}=[(1,2) \to (2,1)]$, which factorizes in a single cyclic permutation of length two, $\mbox{\it ex} = (1 \to 2 \to 1)$. Hence the cycle structure of $P^{-1} P'$ is $\{2\}$, and $V_{{ex},{id}} = V_{2}$. Treating the remaining two terms of Eq.\ (\ref{eq:UpermEx1}) similarly, we obtain
\begin{eqnarray} \label{eq:UpermEx2}
  \left\langle U^{\vphantom{*}}_{a_1 b_1} U^{\vphantom{*}}_{a_2 b_2} 
    U^{*}_{\alpha_1 \beta_1} U^{*}_{\alpha_2 \beta_2} \right\rangle &=& 
  V_{1,1}\, \delta_{a_1 \alpha_1} \delta_{b_1 \beta_1}
          \delta_{a_2 \alpha_2} \delta_{b_2 \beta_2} +
  V_{2}\,   \delta_{a_1 \alpha_2} \delta_{b_1 \beta_1}
          \delta_{a_2 \alpha_1} \delta_{b_2 \beta_2} \nonumber \\ && \mbox{} +
  V_{2}\,   \delta_{a_1 \alpha_1} \delta_{b_1 \beta_2}
          \delta_{a_2 \alpha_2} \delta_{b_2 \beta_1} +
  V_{1,1}\, \delta_{a_1 \alpha_2} \delta_{b_1 \beta_2}
          \delta_{a_2 \alpha_1} \delta_{b_2 \beta_1}.
\end{eqnarray}
In general, the coefficient $V_{1,\ldots,1}$ refers to equal permutations $P=P'$, corresponding to a pairwise (Gaussian) contraction of the matrices $U$ and $U^{*}$. Coefficients $V_{c_1,\ldots,c_k}$ with some $c_j \neq 1$ give non-Gaussian contributions.

The coefficients $V$ are determined by the recursion relation \cite{Samuel}
\begin{eqnarray} \label{eq:Veq}
 N V_{c_1,\ldots,c_k} + \sum_{p+q=c_1} V_{p,q,c_2,\ldots,c_k} + \sum_{j=2}^{k} c_j\, V_{c_1 + c_j,c_2,\ldots,c_{j-1},c_{j+1},\ldots,c_{k}} = \delta_{c_1 1} V_{c_2,\ldots,c_k},
\end{eqnarray}
with $V_0 \equiv 1$. One can show that the solution $V_{c_1,\ldots,c_k}$ does not depend on the order of the indices $c_1,\ldots,c_k$. Results for $V$ up to $n=5$ are given in App.\ \ref{app:coeff}. The large-$N$ expansion of $V$ is
\begin{mathletters} \label{eq:VNlarge}
\begin{eqnarray}
  && V_{c_1,\ldots,c_k} = \prod_{j=1}^{k} V_{c_j} + {\cal O}(N^{k-2n-2}), \\
  && V_{c} = {1 \over c} N^{1-2 c} (-1)^{c-1} {2 c - 2 \choose c-1} + {\cal O}(N^{-1-2c}).
\end{eqnarray}
\end{mathletters}%
(The numbers $c^{-1} {2c-2 \choose c-1}$ are the Catalan numbers.) For example, $V_{1,\ldots,1} = N^{-n} + {\cal O}(N^{-n-2})$. The Gaussian approximation amounts to setting all $V$'s equal to zero except $V_{1,\ldots,1}$, which is set to $N^{-n}$.

The coefficients $V_{c_1,\ldots,c_k}$ determine the moments of $U$. Similarly, the coefficients $W_{c_1,\ldots,c_k}$ determine the cumulants of $U$. The cumulants are obtained from the moments by  subsequent subtraction of all possible factorizations in cumulants of lower degree. For example,
\begin{mathletters}
\begin{eqnarray}
  W_{c_1} &=& V_{c_1}, \\
  W_{c_1,c_2} &=& V_{c_1,c_2} - W_{c_1} W_{c_2}, \\
  W_{c_1,c_2,c_3} &=& V_{c_1,c_2,c_3} - W_{c_1} W_{c_2,c_3} - W_{c_2} W_{c_1,c_3} - W_{c_3} W_{c_1,c_2} - W_{c_1} W_{c_2} W_{c_3}.
\end{eqnarray}
\end{mathletters}%
The recursion relation (\ref{eq:Veq}) for $V$ implies a recursion relation for $W$,
\begin{eqnarray} \label{eq:Weq}
  N W_{c_1,\ldots,c_k} + \sum_{p+q=c_1} W_{p,q,c_2,\ldots,c_k} + \sum_{j=2}^{k} c_j\, W_{c_1 + c_j,c_2,\ldots,c_{j-1},c_{j+1},\ldots,c_{k}} && \nonumber \\ \mbox{} +
  \sum_{p+q=c_1} \sum_{l=1}^{k} {1 \over (l-1)! (k-l)!} \sum_{P} W_{p,c_{P(2)},\ldots,c_{P(l)}} W_{q,c_{P(l+1)},c_{P(k)}} &=& 0,
\end{eqnarray}
with $W_{0} \equiv 1$ and $P$ a permutation of $2,\ldots,k$. To leading order in $1/N$ this equation has the solution,
\begin{eqnarray} \label{eq:WNlarge}
  W_{c_1,\ldots,c_k} = 2^k N^{-2n-k+2} (-1)^{n+k} {(2n + k - 3)! \over (2n)!} \prod_{j=1}^{k} {(2 c_j-1)! \over (c_j-1)!^2} + {\cal O}(N^{-2n-k}).
\end{eqnarray}
Notice that $W_{c_1,\ldots,c_k}$ decreases with increasing number of cycles $k$, opposite to the behavior of $V_{c_1,\ldots,c_k}$. 

In principle, the recursion relations permit an exact evaluation of the average of any polynomial function of $U$. In practice, as the number of $U$'s and $U^{*}$'s increases, keeping track of the indices and of the Kronecker delta's which connect them becomes more and more cumbersome. It is by the introduction of a diagrammatic technique that one can carry out this bookkeeping problem in a controlled and systematic way.

\section{Diagrammatic technique} \label{sec:DIAGR}

The usefulness of diagrams for the bookkeeping problem is well-established for averages over the Gaussian ensemble of Hermitian matrices \cite{Pandey}. Br\'ezin and Zee \cite{BrezinZee} have developed a diagrammatic method which can be applied to non-Gaussian ensembles as well, as a perturbation expansion in a small parameter multiplying the non-Gaussian terms in the distribution. No such small parameter exists for the circular ensemble. The method presented here deals with non-Gaussian contributions to all orders. Creutz \cite{Creutz} has given a diagrammatic algorithm for integrals over SU$(N)$. Because of the more complicated structure of SU$(N)$, we could not effectively apply his method to integrals over ${\cal U}(N)$ in the case of a large number of $U$'s.

\begin{figure}
\hspace{0.35\hsize}
\epsfxsize=0.3 \hsize
\epsffile{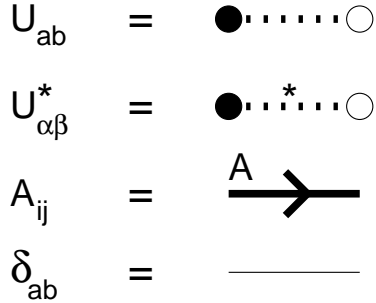}
\medskip

\caption{\label{fig:UAVG1} Substitution rules for the unitary matrices $U$ and $U^{*}$, the fixed matrix $A$ and the Kronecker delta.}
\end{figure}

\begin{figure}
\hspace{0.3\hsize}
\epsfxsize=0.4 \hsize
\epsffile{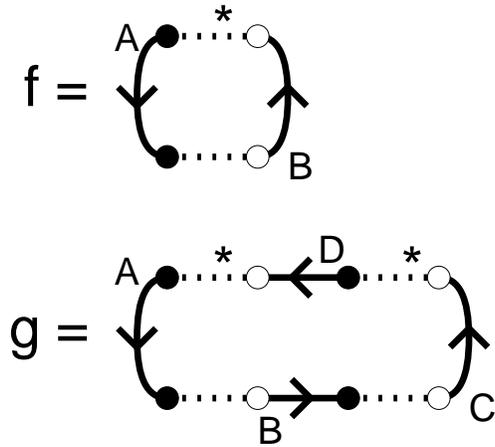}
\medskip

\caption{\label{fig:UAVG3} Diagrammatic representation of the functions $f(U) = \mbox{tr}\, A U B U^{\dagger}$ and $g(U) = \mbox{tr}\, A U B U C U^{\dagger} D U^{\dagger}$.}
\end{figure}

The diagrams consist of the building blocks shown in Fig.\ \ref{fig:UAVG1}. We represent matrix elements $U_{ab}^{\vphantom{*}}$ or $U^{*}_{\alpha\beta}$ by thick dotted lines. The first index ($a$ or $\alpha$) is a black dot, the second index ($b$ or $\beta$) a white dot. A fixed matrix $A_{ij}$ is represented by a directed thick solid line, pointed from the first to the second index. Summation over an index is indicated by attachment of the solid line to a dot. As an example, the functions $f(U) = \mbox{tr}\, A U B U^{\dagger}$ and $g(U) = \mbox{tr}\, A U B U C U^{\dagger} D U^{\dagger}$ are represented in Fig.\ \ref{fig:UAVG3}.

The average over the matrix $U$ consists of summing over all permutations $P$ and $P'$ in Eq.\ (\ref{eq:Uperm}). Permutations are generated by drawing thin lines (representing Kronecker deltas) between all black dots attached to $U$ and black dots attached to $U^{*}$ (one line per dot). Black dots connect to black dots and white dots to white dots. To find the contribution of the permutations $P$ and $P'$ to $\langle f(U) \rangle$, we need (i) to determine the cycle structure of the permutation $P^{-1} P'$, and (ii) to sum over the indices of the fixed matrices $A$. 

(i) The cycle structure can be read off from the diagrams. A cycle of the permutation $P^{-1}P'$ gives rise to a closed circuit in the diagram consisting of alternating dotted and thin lines. The length $c_k$ of the cycle is half the number of dotted lines contained in the circuit. We call such circuits $U$-cycles of length $c_k$.

(ii) The trace over the elements of $A$ is done by inspection of the closed circuits in the diagram which consist of alternating thick and thin lines. We call such circuits $T$-cycles. A $T$-cycle containing the matrices $A^{(1)}$, $A^{(2)}$, \ldots, $A^{(k)}$ (in this order) gives rise to $\mbox{tr}\, A^{(1)} A^{(2)} \ldots A^{(k)}$. If the thick line corresponding to a matrix $A$ is traversed opposite to its direction, the matrix should be replaced by its transpose $A^{\rm T}$.

\begin{figure}
\hspace{0.12\hsize}
\epsfxsize=0.75 \hsize
\epsffile{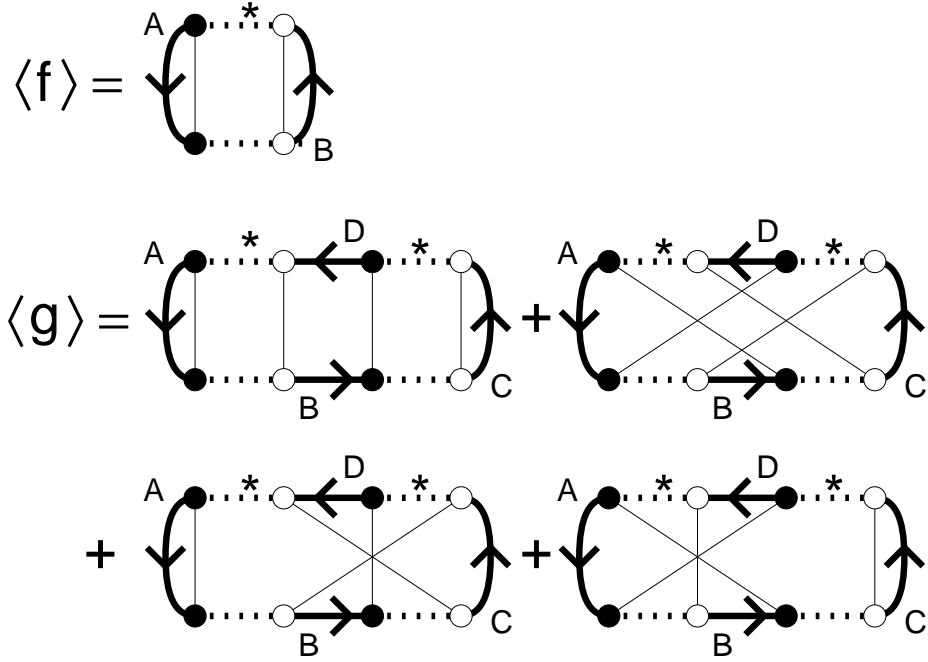}
\medskip

\caption{\label{fig:UAVG4} Diagrammatic representation of the averages of the functions $f$ and $g$ in Fig.\ \protect\ref{fig:UAVG3}.}
\end{figure}

As an example, let us consider the average of the functions $f(U) = \mbox{tr}\, A U B U^{\dagger}$ and $g(U) = \mbox{tr}\, A U B U C U^{\dagger} D U^{\dagger}$. Connecting the dots by thin lines, we arrive at the diagrams of Fig.\ \ref{fig:UAVG4}. For $f$, there is only one diagram. It contains a single $U$-cycle of length $1$ (weight $V_{1}$) and two $T$-cycles (which generate $\mbox{tr}\, A$ and $\mbox{tr}\, B$). We look up the value of $V_{1} = 1/N$ in App.\ \ref{app:coeff}, and find
\begin{equation} \label{eq:exF}
  \langle f(U) \rangle = V_{1} \mbox{tr}\, A\, \mbox{tr}\, B = N^{-1} \mbox{tr}\, A\, \mbox{tr}\, B,
\end{equation}
Four diagrams contribute to $g$. The first diagram contains two $U$-cycles of length $1$, and three $T$-cycles. Its contribution is $V_{1,1}\, \mbox{tr}\, A\, \mbox{tr}\, B D\, \mbox{tr}\, C$. The second diagram contains two $U$-cycles of length $1$ and a single $T$-cycle. Its contribution is $V_{1,1} \mbox{tr}\, A D C B$. The third and fourth diagram each contain a single $U$-cycle of length $2$ and two $T$-cycles. Their contributions are $V_{2}\, \mbox{tr}\, A\, \mbox{tr}\, B D C$ and $V_{2}\, \mbox{tr}\, A D B\, \mbox{tr}\, C$. In total we find
\begin{eqnarray}
  \langle g(U) \rangle &=& V_{1,1} (\mbox{tr}\, A\, \mbox{tr}\, B D\, \mbox{tr}\, C\, + \mbox{tr}\, A D C B\,) + V_{2} (\mbox{tr}\, A\, \mbox{tr}\, B D C\, + \mbox{tr}\, A D B\, \mbox{tr}\, C) \nonumber \\ &=&
  (N^2-1)^{-1} \left( \mbox{tr}\, A\, \mbox{tr}\, BD\, \mbox{tr}\, C \, + \mbox{tr}\, A D C B \right) \nonumber \\ && \mbox{} - [N(N^2-1)]^{-1} \left( \mbox{tr}\, A\, \mbox{tr}\, B D C + \mbox{tr}\, A D B\, \mbox{tr}\, C \right). \label{eq:exG}
\end{eqnarray}

Whereas each individual $T$-cycle gives rise to a trace of matrices, it is only the combination of {\em all} $U$-cycles together that determines the coefficient $V_{c_1,\ldots,c_k}$. The evaluation of a diagram would be more efficient, if we could attribute a weight to an {\em individual} $U$-cycle. We introduced the cumulant expansion of the coefficients $V$ in the coefficients $W$ for this purpose. The leading term $V_{c_1,\ldots,c_k} = \prod_{p=1}^{k} W_{c_p}$ of the cumulant expansion attributes a weight $W_{c_p}$ to each individual $U$-cycle of length $c_p$. This is sufficient for the calculation of the large-$N$ limit of the average $\langle f \rangle $. The next term $\sum_{i < j}^{k} W_{c_i,c_j} \prod_{p \neq i,j}^{k} W_{c_p}$ attributes a weight $W_{c_i,c_j}$ to the pair $(i,j)$ of $U$-cycles, and the weight $W_{c_p}$ to all others individually. This is sufficient for the variance of $f$. The general rule is that the $j$th order cumulant of $f$ in the large-$N$ limit requires the $j$th order term in the cumulant expansion of the coefficients $V$, and hence requires consideration of groups of $j$ $U$-cycles.

Let us summarize the diagrammatic rules:
\begin{enumerate}
\item Draw the diagrams according to the substitution rules of Fig.\ \ref{fig:UAVG1}.
\item Draw thin lines to pair black dots attached to $U$ to black dots attached to $U^{*}$. Do the same for the white dots.
\item Every closed circuit of alternating thick solid lines and thin solid lines (a $T$-cycle) corresponds to a trace of the matrices $A$ appearing in the circuit. If a thick line  is traversed opposite to its direction, the transpose of the matrix appears in the trace.
\item Every closed circuit of alternating dotted and thin solid lines (a $U$-cycle) corresponds to a cycle of length $c_k$ equal to half the number of dotted lines. The set of $U$-cycles in a diagram defines the coefficient $V_{c_1,\ldots,c_k}$, which is the weight of the diagram. The coefficient $V$ can be factorized into cumulants. To determine the cumulant coefficients $W$, partition the $U$-cycles into groups. Every group of $p$ $U$-cycles of lengths $c_1$, \ldots, $c_p$ contributes a weight $W_{c_1,\ldots,c_p}$. 
\end{enumerate}

The diagrammatic rules are exact. In the large-$N$ limit, we may reduce the number of diagrams and partitions that is involved. Let us determine the order in $N$ of a diagram with $l$ $T$-cycles and $k$ $U$-cycles of total length $n$ partitioned into $g$ groups. Counting every trace as an order $N$ and using the large-$N$ result (\ref{eq:WNlarge}) for the coefficients $W$, we find a contribution of order $N^{2g+l-k-2n}$. Since $g \le k$ the order is maximal if $g=k$ and the total number of cycles $k+l$ is maximal. Thus, for large $N$, we may restrict ourselves to diagrams with as many cycles as possible and with a partition of the $U$-cycles in groups of a single cycle (i.e.\ we may approximate $V_{c_1,\ldots,c_k} \approx W_{c_1} \ldots W_{c_k}$). 

\begin{figure}
\hspace{0.1\hsize}
\epsfxsize=0.8 \hsize
\epsffile{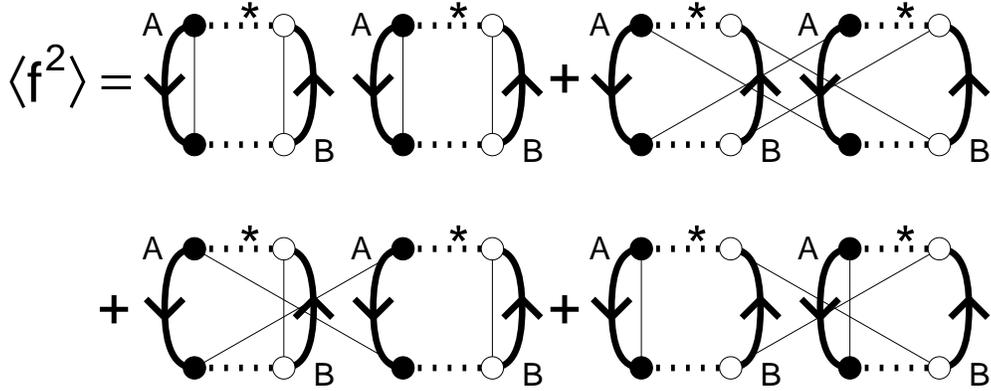}
\medskip

\caption{\label{fig:UAVG5} Diagrammatic representation of $\langle f^2 \rangle$.}
\end{figure}%

We conclude this section with one more example, which is the calculation of the variance $\mbox{var}\, f = \langle f^2 \rangle - \langle f \rangle^2$ of the function $f(U) = \mbox{tr}\, A U B U^{\dagger}$. Diagrammatically, we calculate $\langle f^2 \rangle$ as in Fig.\ \ref{fig:UAVG5}a, resulting in
\begin{mathletters}\label{eq:exvariance}
\begin{eqnarray} 
  \langle f^2 \rangle &=& V_{1,1}\, \left[ (\mbox{tr}\, A)^2\, (\mbox{tr}\, B)^2 + \mbox{tr}\, A^2\, \mbox{tr}\, B^2 \right] + W_{2} \left[ \mbox{tr}\, A^2 (\mbox{tr}\, B)^2 + (\mbox{tr}\, A)^2\, \mbox{tr}\, B^2 \right], \\
  \Longrightarrow\ \mbox{var}\, f &=& W_{1,1}\, \left[ (\mbox{tr}\, A)^2\, (\mbox{tr}\, B)^2 + \mbox{tr}\, A^2\, \mbox{tr}\, B^2 \right] + W_{1}^{2}\, \mbox{tr}\, A^2\, \mbox{tr}\, B^2 \nonumber \\ && \mbox{} + W_{2} \left[ \mbox{tr}\, A^2 (\mbox{tr}\, B)^2 + (\mbox{tr}\, A)^2\, \mbox{tr}\, B^2 \right].
\end{eqnarray}
\end{mathletters}%
If we now consider the order in $N$ of the various contributions, we see that the leading ${\cal O}(N^2)$ term of $\langle f^2 \rangle$ ($l=4$, $g=k=2$, corresponding to $6$ cycles and a partition of the $U$-cycles into two groups of a single cycle), is exactly canceled by $\langle f \rangle^2$. This exact cancelation is possible because the leading contribution of $\langle f^2 \rangle$ is {\em disconnected}: Each $T$-cycle, and each group of $U$-cycles belongs entirely to one of the two factors $\mbox{tr}\, A U B U^{\dagger}$ of $f^2$. Only connected diagrams contribute to the variance of $f$. The connected diagrams are of order $1$ ($k+l = 4$ and $g=k$ or $k+l=6$ and $g=k-1$). They give the variance
\begin{eqnarray}
  \mbox{var}\, f &=& W_{1,1} (\mbox{tr}\, A)^2\, (\mbox{tr}\, B)^2 + W_{1}^2 \mbox{tr}\, A^2\, \mbox{tr}\, B^2 \nonumber \\ && \mbox{} + W_{2} \left[ \mbox{tr}\, A^2\, (\mbox{tr}\, B)^2 + (\mbox{tr}\, A)^2\, \mbox{tr}\, B^2 \right] + {\cal O}(N^{-1}).
\end{eqnarray}

\section{Integration of unitary symmetric matrices} \label{sec:UAVGb1}

In the presence of time-reversal symmetry the scattering matrix $S$ is both unitary and symmetric: $S S^{\dagger} = 1$, $S = S^{\rm T}$. The elements of $S$ are complex numbers. (The case of a quaternion $S$, corresponding to spin-orbit scattering, is treated in the next section.) The ensemble of uniformly distributed unitary symmetric matrices is known as the circular orthogonal ensemble (COE) \cite{Dyson,Mehta}. Averages of the unitary symmetric matrix $U$ over the COE can be computed in two ways. One way is to substitute $U = V V^{\rm T}$, with the matrix $V$ uniformly distributed over the unitary group. This has the advantage that one can use the same formulas as for averages over the CUE, but the disadvantage that the number of unitary matrices is doubled. A more efficient way is to use specific formulas for the COE, as we now discuss.

The average of a polynomial in $U$ and $U^{*}$ over the COE has the general structure
\begin{equation} \label{eq:Uparamb1} \label{eq:Uavgb1}
  \langle U^{\vphantom{*}}_{a_1 a_2} \ldots U^{\vphantom{*}}_{a_{2n-1} a_{2n}} U^{*}_{\alpha_{1} \alpha_{2}} \ldots U^{*}_{\alpha_{2m-1} \alpha_{2m}} \rangle = \delta_{n m} \sum_{P} V_{P} \prod_{j=1}^{2n} \delta_{a_j \alpha_{P(j)}}.
\end{equation}
The summation is over permutations $P$ of the numbers $1$, \ldots, $2n$. We can decompose $P$ as
\begin{equation}
  P = \left(\prod_{j=1}^{n} T_j\right) P_{\rm e} P_{\rm o} \left(\prod_{j=1}^{n} T'_j\right),
\end{equation}
where $T_j$ and $T_j'$ permute the numbers $2j-1$ and $2j$, and $P_{\rm e}$ ($P_{\rm o}$) permutes $n$ even (odd) numbers. Because $U_{ab} = U_{ba}$, the moment coefficient $V_{P}$ depends only on the cycle structure $\{ c_1,\ldots,c_k \}$ of $P_{\rm e}^{-1} P_{\rm o}^{\vphantom{-1}}$ \cite{MelloSeligman}, so that we may write $V_{c_1,\ldots,c_k}$ instead of $V_{P}$.

The moment coefficients obey the recursion relation
\begin{eqnarray} \label{eq:Veqb1}
  (N + c_1) V_{c_1,\ldots,c_k} + \sum_{p+q=c_1} V_{p,q,c_2,\ldots,c_k} + 2 \sum_{j=2}^{k} c_j\, V_{c_1 + c_j,c_2,\ldots,c_{j-1},c_{j+1},\ldots,c_{k}} = \delta_{c_1 1} V_{c_2,\ldots,c_k},
\end{eqnarray}
with $V_{0} \equiv 1$. The large-$N$ expansion of $V$ is
\begin{mathletters} \label{eq:VNlargeb1}
\begin{eqnarray} 
  && V_{c_1,\ldots,c_k} = \prod_{j=1}^{k} V_{c_{j}} + {\cal O}(N^{k-2n-2}), \\
  && V_{c} = {1 \over c} N^{1-2c} (-1)^{c-1} {2c-2 \choose c - 1} - N^{-2c} (-4)^{c-1} + {\cal O}(N^{-1-2c}).
\end{eqnarray}
\end{mathletters}%
Compared with Eq.\ (\ref{eq:VNlarge}) an extra term of order $N^{-2c}$ appears in $V_{c}$ because of the symmetry restriction. The recursion relation for the cumulant coefficients $W$ is
\begin{eqnarray} \label{eq:Weqb1}
  (N + c_1) W_{c_1,\ldots,c_k} + \sum_{p+q=c_1} W_{p,q,c_2,\ldots,c_k} + 2 \sum_{j=2}^{k} c_j\, W_{c_1 + c_j,c_2,\ldots,c_{j-1},c_{j+1},\ldots,c_{k}} + && \nonumber \\ \mbox{} +
  \sum_{p+q=c_1} \sum_{l=1}^{k} {1 \over (l-1)! (k-l)!} \sum_{P} W_{p,c_{P(2)},\ldots,c_{P(l)}} W_{q,c_{P(l+1)},c_{P(k)}} &=& 0,
\end{eqnarray}
with $W_0 \equiv 1$ and $P$ a permutation of the numbers $2,\ldots,k$. The solution for large $N$ is
\begin{equation}
  W_{c_1,\ldots,c_k} = 2^{2k-1} N^{-2n-k+2} (-1)^{n+k} {(2n + k - 3)! \over (2n)!} \prod_{j=1}^{k} {(2 c_j-1)! \over (c_j-1)!^2} + {\cal O}(N^{-2n-k+1}).
\end{equation}
The coefficients $V_{c_1,\ldots,c_k}$ and $W_{c_1,\ldots,c_k}$ are listed in App.\ \ref{app:coeff} for  $n = c_1 + \ldots + c_k \le 5$.

For the diagrammatic representation, we again use the substitution rules of Fig.\ \ref{fig:UAVG1}. The symmetry of $U$ is taken into account by allowing thin lines between black and white dots. Therefore, rule 2 is replaced by
\begin{itemize}
\item[2.] Pair the dots attached to $U$ to the dots attached to $U^{*}$ by connecting them with thin lines.
\end{itemize}

As examples, we compute the averages of $f(U) = \mbox{tr}\, A U B U^{\dagger}$ and $g(U) = \mbox{tr}\, A U B U C U^{\dagger} D U^{\dagger}$ over the COE. The diagrams for $\langle f(U) \rangle$ are shown in Fig.\ \ref{fig:UAVG7}, with the result
\begin{mathletters} \label{eq:FGexb1}
\begin{eqnarray} \label{eq:Fexb1}
  \langle f(U) \rangle = V_{1} (\mbox{tr}\, A\, \mbox{tr}\, B + \mbox{tr}\, A^{\rm T} B) = (N+1)^{-1} (\mbox{tr}\, A\, \mbox{tr}\, B + \mbox{tr}\, A^{\rm T} B).
\end{eqnarray}
\begin{figure}
\hspace{0.25\hsize}
\epsfxsize=0.5 \hsize
\epsffile{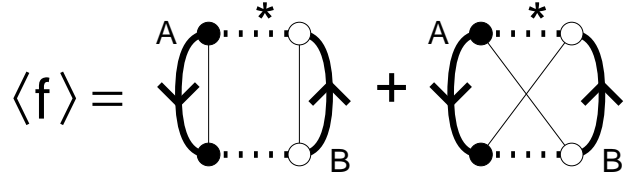}
\medskip

\caption{\label{fig:UAVG7} Diagrammatic representation of $\langle f(U) \rangle$ for $f(U) = \mbox{tr}\, A U B U^{\dagger}$, where $U$ is a unitary symmetric matrix. The second term arises because of the symmetry constraint.}
\end{figure}%
Similarly, we find that
\begin{eqnarray} 
  \langle g(U) \rangle &=& 
[(N+1)(N+3)]^{-1} \nonumber \\ && \mbox{} \times (\mbox{tr}\, A\, \mbox{tr}\, B D\, \mbox{tr}\, C\, + \mbox{tr}\, A D^{\rm T} B^{\rm T}\, \mbox{tr}\, C\, + \mbox{tr}\, A\, \mbox{tr}\, B C^{\rm T} D\, + \mbox{tr}\, A D^{\rm T} C B^{\rm T}   \nonumber \\ && \mbox{} + \mbox{tr}\, A D C B\, + \mbox{tr}\, A C^{\rm T} D^{\rm T} B + \mbox{tr}\, A D B^{\rm T} C^{\rm T} + \mbox{tr}\, A C^{\rm T}\, \mbox{tr}\, B D^{\rm T}) \nonumber \\ && \mbox{} - [(N(N+1)(N+3)]^{-1} \nonumber \\ && \mbox{} \times (\mbox{tr}\, A\, \mbox{tr}\, B D C\, + \mbox{tr}\, A C^{\rm T} D^{\rm T} B^{\rm T}\, + \mbox{tr}\, A\, \mbox{tr}\, B D^{\rm T} C\, + \mbox{tr}\, A C^{\rm T} D B^{\rm T}  \nonumber \\ && \mbox{} + \mbox{tr}\, A D B\, \mbox{tr}\, C + \mbox{tr}\, A D^{\rm T} B\, \mbox{tr}\, C\, + \mbox{tr}\, A D B C^{\rm T} + \mbox{tr}\, A D^{\rm T} B C^{\rm T} \nonumber \\ && \mbox{} + \mbox{tr}\, A D^{\rm T} B^{\rm T} C^{\rm T}\, + \mbox{tr}\, A C^{\rm T}\, \mbox{tr}\, B D\, + \mbox{tr}\, A D^{\rm T} C B\, + \mbox{tr}\, A C^{\rm T} D B \nonumber \\ && \mbox{} + \mbox{tr}\, A\, \mbox{tr}\, B C^{\rm T} D^{\rm T}\, + \mbox{tr}\, A D C B^{\rm T}\, + \mbox{tr}\, A\, \mbox{tr}\, B D^{\rm T}\, \mbox{tr}\, C\, + \mbox{tr}\, A D B^{\rm T}\, \mbox{tr}\, C).
\end{eqnarray}
\end{mathletters}%

\section{Integration of matrices of quaternions} \label{sec:trans}

We extend the results of the previous sections for integrals over unitary matrices of complex numbers to integrals over unitary matrices of quaternions. This is relevant to the case that spin-rotation symmetry is broken by spin-orbit scattering.

Let us first recall the definition and basic properties of quaternions \cite{Mehta}. A quaternion $q$ is represented by a $2 \times 2$ matrix,
\begin{equation}
  q = a_{0} \openone + i a_{1} \sigma_1 + i a_{2} \sigma_2 + i a_{3} \sigma_3,
\end{equation}
where $\openone$ is the $2 \times 2$ unit matrix and $\sigma_i$ is a Pauli matrix,
\begin{equation}
  \sigma_{1} = \left( \begin{array}{cc} 0 & 1 \\ 1 & 0 \end{array} \right), \ \
  \sigma_{2} = \left( \begin{array}{cc} 0 &-i \\ i & 0 \end{array} \right), \ \
  \sigma_{3} = \left( \begin{array}{cc} 1 & 0 \\ 0 &-1 \end{array} \right).
\end{equation}
The coefficients $a_{j}$ are complex numbers. The complex conjugate $q^{*}$ and Hermitian conjugate $q^{\dagger}$ of a quaternion $q$ are defined as
\begin{equation}
  q^{*} = a_{0}^{*} \openone + i a_{1}^{*} \sigma_1 + i a_{2}^{*} \sigma_2 + i a_{3}^{*} \sigma_3,\ \
  q^{\dagger} = a_{0}^{*} \openone - i a_{1}^{*} \sigma_1 - i a_{2}^{*} \sigma_2 - i a_{3}^{*} \sigma_3.
\end{equation}
The complex conjugate of a quaternion differs from the complex conjugate of a $2 \times 2$ matrix, whereas the Hermitian conjugate equals the Hermitian conjugate of a $2 \times 2$ matrix. 
Let $Q$ be an $N \times N$ matrix of quaternions with elements $Q_{kl} = Q_{kl}^{(0)} \openone + i Q_{kl}^{(1)} \sigma_1 + i Q_{kl}^{(2)} \sigma_2 + i Q_{kl}^{(3)} \sigma_3$. The complex conjugate $Q^{*}$ and Hermitian conjugate $Q^{\dagger}$ are defined by $(Q^{*})_{kl} = Q_{kl}^{*}$ and $(Q^{\dagger})_{kl} = Q_{lk}^{\dagger}$. The dual matrix $Q^{\rm R}$ is defined by $Q^{\rm R} = (Q^{\dagger})^{*} = (Q^{*})^{\dagger}$. We call $Q$ unitary if $Q Q^{\dagger} = 1$ and self-dual if $Q = Q^{\rm R}$. A unitary self-dual matrix is defined by $Q Q^{\dagger} = Q Q^{*} = 1$. The trace $\mbox{tr}\, Q$ is defined by $\mbox{tr}\, Q = \sum_{j} Q_{jj}^{(0)}$, which equals $1/2$ the trace of the $2N \times 2N$ complex matrix corresponding to $Q$. The scattering matrix in zero magnetic field is a unitary self-dual matrix, because of time-reversal symmetry. The ensemble of quaternion matrices which is uniformly distributed over the unitary group is called the circular unitary ensemble (CUE). If the ensemble is restricted to self-dual matrices it is called the circular symplectic ensemble (CSE) \cite{Dyson,Mehta}.

The integration of a polynomial function $f(U)$ of an $N \times N$ quaternion matrix $U$ over the CUE or CSE can be related to the integration of a function ${\hat f}(U)$ of an $N \times N$ complex matrix $U$ over the CUE or COE. The translation rule is as follows (a similar rule has been formulated for Gaussian ensembles in Refs.\ \onlinecite{Juengling,Wegner}): 
\begin{enumerate}
\item  ${\hat f}(U)$ is constructed from $f(U)$ by replacing, respectively, the complex conjugates, Hermitian conjugates, and duals of quaternion matrices by complex conjugates, Hermitian conjugates, and transposes of complex matrices. Furthermore, every trace is replaced by $-\case{1}{2} \mbox{tr}\, $, and numerical factors $N$ are replaced by $-\case{1}{2}N$. 
\item  The average $\langle {\hat f}(U) \rangle$ is calculated using the rules for integration of $N \times N$ complex matrices over the CUE or COE.
\item  The average $\langle f(U) \rangle$ over the CUE or CSE is found by replacing, respectively,  the complex conjugates, Hermitian conjugates, and transposes of complex matrices by the complex conjugates, Hermitian conjugates, and duals of quaternion matrices. Traces are replaced by $-2\, \mbox{tr}\, $ and numerical factors $N$ by $-2 N$.
\end{enumerate}
                          
As examples, we compute the averages of the functions $f(U) = \mbox{tr}\, A U B U^{\dagger}$ and $g(U) = \mbox{tr}\, A U B U C U^{\dagger} D U^{\dagger}$ of $N \times N$ quaternion matrices over the CUE and CSE. The first step is to construct the functions ${\hat f}(U)$ and ${\hat g}(U)$ of $N \times N$ complex matrices,
\begin{equation}
  {\hat f}(U) = -\case{1}{2}\, \mbox{tr} A U B U^{\dagger}, \ \
  {\hat g}(U) = -\case{1}{2}\, \mbox{tr} A U B U C U^{\dagger} D U^{\dagger}.
\end{equation}
The second step is to average ${\hat f}$ and ${\hat g}$ over the CUE. The result is in Eqs.\ (\ref{eq:exF}) and (\ref{eq:exG}),
\begin{mathletters}
\begin{eqnarray}
  \langle {\hat f} \rangle_{\rm CUE} &=& -\case{1}{2} N^{-1} \mbox{tr}\, A\, \mbox{tr}\, B, \\
  \langle {\hat g} \rangle_{\rm CUE} &=& -\case{1}{2}
  (N^2-1)^{-1} \left( \mbox{tr}\, A\, \mbox{tr}\, BD\, \mbox{tr}\, C \, + \mbox{tr}\, A D C B \right) \nonumber \\ && \mbox{} + \case{1}{2} [N(N^2-1)]^{-1} \left( \mbox{tr}\, A\, \mbox{tr}\, B D C + \mbox{tr}\, A D B\, \mbox{tr}\, C \right).
\end{eqnarray}
\end{mathletters}%
The third step is to translate back to quaternion matrices,
\begin{mathletters}
\begin{eqnarray}
  \langle {f} \rangle_{\rm CUE} &=& N^{-1} \mbox{tr}\, A\, \mbox{tr}\, B, \\
  \langle {g} \rangle_{\rm CUE} &=& 
  (4 N^2-1)^{-1} \left( 4\, \mbox{tr}\, A\, \mbox{tr}\, BD\, \mbox{tr}\, C \, + \mbox{tr}\, A D C B \right) \nonumber \\ && \mbox{} - [N(4 N^2-1)]^{-1} \left( \mbox{tr}\, A\, \mbox{tr}\, B D C + \mbox{tr}\, A D B\, \mbox{tr}\, C \right).
\end{eqnarray}
\end{mathletters}%
Similarly, to average of $f$ and $g$ over the CSE we need the average of ${\hat f}$ and ${\hat g}$ over the COE given by Eq.\ (\ref{eq:Fexb1}), and then translate back to quaternion matrices. For $\langle f(U) \rangle$ we find
\begin{mathletters}
\begin{eqnarray}
  \langle {\hat f} \rangle_{\rm COE} &=& -\case{1}{2} (N+1)^{-1} (\mbox{tr}\, A\, \mbox{tr}\, B + \mbox{tr}\, A^{\rm T} B), \nonumber \\
  \Longrightarrow 
  \langle f \rangle_{\rm CSE} &=& (2N-1)^{-1} (2\, \mbox{tr}\, A\, \mbox{tr}\, B - \mbox{tr}\, A^{\rm R} B).
\end{eqnarray}
Similarly, we find for $\langle g(U) \rangle$ the final result
\begin{eqnarray}
  \langle g \rangle_{\rm CSE} &=&
  [(2N-1)(2N-3)]^{-1}
  \nonumber \\ && \mbox{}
  \times (4\, \mbox{tr}\, A\, \mbox{tr}\, B D\, \mbox{tr}\, C\,
    - 2\, \mbox{tr}\, A D^{\rm R} B^{\rm R}\, \mbox{tr}\, C\,
    - 2\,  \mbox{tr}\, A\, \mbox{tr}\, B C^{\rm R} D\,
    + \mbox{tr}\, A D^{\rm R} C B^{\rm R} 
  \nonumber \\ && \mbox{}
    + \mbox{tr}\, A D C B\,
    + \mbox{tr}\, A C^{\rm R} D^{\rm R} B
    + \mbox{tr}\, A D B^{\rm R} C^{\rm R}
    - 2\, \mbox{tr}\, A C^{\rm R}\, \mbox{tr}\, B D^{\rm R})
  \nonumber \\ && \mbox{}
    - [(2N(2N-1)(2N-3)]^{-1}
  \nonumber \\ && \mbox{}
  \times (2\, \mbox{tr}\, A\, \mbox{tr}\, B D C\,
    - \mbox{tr}\, A C^{\rm R} D^{\rm R} B^{\rm R}\,
    + 2\, \mbox{tr}\, A\, \mbox{tr}\, B D^{\rm R} C\,
    - \mbox{tr}\, A C^{\rm R} D B^{\rm R}
  \nonumber \\ && \mbox{}
    + 2\, \mbox{tr}\, A D B\, \mbox{tr}\, C
    + 2 \, \mbox{tr}\, A D^{\rm R} B\, \mbox{tr}\, C\,
    - \mbox{tr}\, A D B C^{\rm R}
    - \mbox{tr}\, A D^{\rm R} B C^{\rm R}
  \nonumber \\ && \mbox{}
    + 2\, \mbox{tr}\, A\, \mbox{tr}\, B C^{\rm R} D^{\rm R}\,
    - \mbox{tr}\, A D C B^{\rm R}\,
    - 4\, \mbox{tr}\, A\, \mbox{tr}\, B D^{\rm R}\, \mbox{tr}\, C\,
    + 2\, \mbox{tr}\, A D B^{\rm R}\, \mbox{tr}\, C
  \nonumber \\ && \mbox{}
    - \mbox{tr}\, A D^{\rm R} B^{\rm R} C^{\rm R}\,
    + 2\, \mbox{tr}\, A C^{\rm R}\, \mbox{tr}\, B D\,
    - \mbox{tr}\, A D^{\rm R} C B\,
    - \mbox{tr}\, A C^{\rm R} D B).
\end{eqnarray}
\end{mathletters}%

\section{Application to a chaotic cavity} \label{sec:QD}

\begin{figure}
\hspace{0.3\hsize}
\epsfxsize=0.4 \hsize
\epsffile{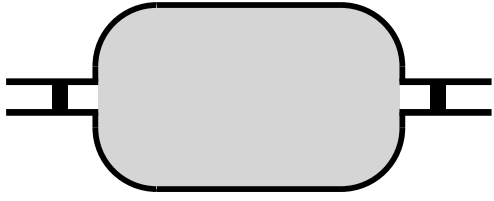}
\medskip

\caption{\label{fig:QD1} Chaotic cavity (grey) connected to two leads containing tunnel barriers (black).}
\end{figure}

We consider the system shown in Fig.\ \ref{fig:QD1}, consisting of a chaotic cavity attached to two leads, containing tunnel barriers. The $M \times M$ scattering matrix $S$ is decomposed into $N_i \times N_j$ submatrices $s_{ij}$,
\begin{equation}
  S = \left( \begin{array}{cc} s_{11} & s_{12} \\ s_{21} & s_{22} \end{array} \right),
\end{equation}
which describe scattering from lead $j$ into lead $i$ ($M = N_i + N_j$). The conductance $G$ is given by the Landauer formula,
\begin{equation} \label{eq:QDlandauer}
  G/G_0 = \mbox{tr}\, s_{12}^{\vphantom{\dagger}} s_{12}^{\dagger} =
          \mbox{tr}\, C_1 S C_2 S^{\dagger},\ \ G_0 = 2e^2/h.
\end{equation}
The projection matrices $C_1$ and $C_2 = 1 - C_1$ are defined by $(C_1)_{ij} = 1$ if $i=j \le N_1$ and $0$ otherwise.

In the absence of tunnel barriers in the leads, $S$ is distributed according to the circular ensemble. The symmetry index $\beta \in \{1,2,4\}$ distinguishes the COE ($\beta=1$), CUE ($\beta=2$), and CSE ($\beta=4$). Calculation of the average and variance of $G$ is straightforward \cite{BarangerMello,JPB},
\begin{eqnarray}
  \langle G/G_0 \rangle &=& 
    { \beta N_1 N_2 \over  \beta M + 2 -\beta}, \label{eq:QDavgC}\\
  \mbox{var}\, G/G_0 &=& 
  {2 \beta N_1 N_2 (\beta N_1 + 2 - \beta)(\beta N_2 + 2 - \beta) \over (\beta M + 2 - 2 \beta)(\beta M + 2 - \beta)^2(\beta M + 4 - \beta)}. \label{eq:QDvarC}
\end{eqnarray}

In the presence of a tunnel barrier in lead $i$ with reflection matrix $r_i$, the distribution of $S$ is given by the Poisson kernel\cite{Hua,MPS,DoronSmilansky,Brouwer},
\begin{equation}
  P(S) \propto
    \left|\det(1 - \bar S^\dagger S)\right|^{-(\beta M + 2 - \beta)}, \ \
   \bar S = \left( \begin{array}{cc} r_1 & 0 \\ 0 & r_2
     \end{array} \right). \label{eq:poisson2}
\end{equation}
The sub-unitary matrix $\bar S$ is the ensemble average of $S$: $\int dS P(S) S = \bar S$. The eigenvalues $\Gamma_j$ of $1 - \bar S \bar S^{\dagger}$ are the transmission eigenvalues of the tunnel barriers. The fluctuating part $\delta S \equiv S - \bar S$ of $S$ can be decomposed as
\begin{equation}
   \label{eq:Sparam}
  \delta S = T'(1- U R')^{-1} U T,
\end{equation}
where $T$, $T'$, and $R'$ are $M \times M$ matrices such that the 
$2M \times 2M$ matrix
\begin{equation} \label{eq:SigmaTdef}
  \Sigma = \left( \begin{array}{cc} \bar S & T' \\ T & R' \end{array} \right)
\end{equation}
is unitary. The usefulness of the decomposition (\ref{eq:Sparam}) is that $U$ is distributed according the circular ensemble \cite{FriedmanMello2,Hua,Brouwer}. In the presence of time-reversal symmetry, we further have $\bar S = \bar S^{\rm T}$, $T' = T^{\rm T}$, $R' = R'^{\rm T}$, and $U = U^{\rm T}$. Physically, $U$ corresponds to the scattering matrix of the cavity without the tunnel barriers in the leads and $\Sigma$ corresponds to the scattering matrix of the tunnel barriers in the absence of the cavity \cite{BrouwerBeenakker94,Brouwer}.

The decomposition (\ref{eq:Sparam}) reduces the problem of averaging $S$ with the Poisson kernel to integrating $U$ over the unitary group. Because the conductance $G$ is a rational function of $U$, this average can not be done in closed form for all $M$. For $N_1,N_2 \gg 1$ a perturbative calculation is possible. In this section we will compute the mean and variance of the conductance in the large-$N$ limit, using the diagrammatic technique of the previous sections.

\subsection{Average conductance} \label{sec:QDavg}

\begin{figure}
\hspace{0.2\hsize}
\epsfxsize=0.6\hsize
\epsffile{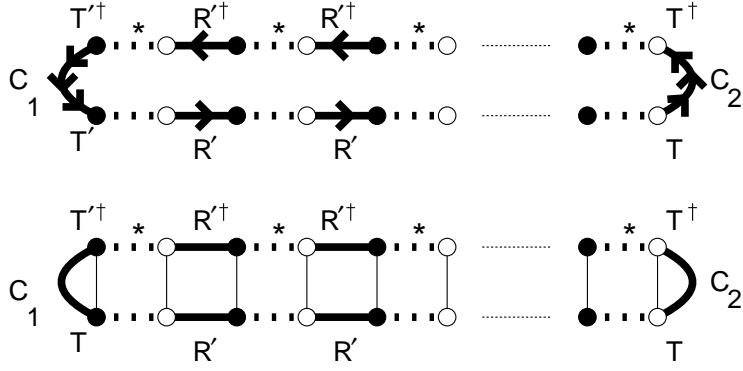}
\medskip

\caption{\label{fig:QDWL1} Top: Diagrammatic representation of the function $f_n(U)$ in Eq.\ (\protect\ref{eq:Gsum}); Bottom: Ladder diagram with the largest number of cycles, which gives the ${\cal O}(N)$ contribution to the average conductance. The arrows are omitted if the direction of the diagram is not ambiguous.}
\end{figure}

According to the Landauer formula (\ref{eq:QDlandauer}) the average conductance is given by
\begin{eqnarray} \label{eq:QDWLcondsum}
  \langle G/G_0 \rangle &=& 
    \langle \mbox{tr}\, C_1 \delta S C_2 \delta S^{\dagger} \rangle,
\end{eqnarray}
where we have used that $\langle \delta S \rangle = 0$. Expansion of the denominator in the decomposition (\ref{eq:Sparam}) of $\delta S$ yields the series
\begin{mathletters} \label{eq:Gsum}
\begin{eqnarray}
    \langle G/G_0 \rangle
  &=& \sum_{n=1}^{\infty} \langle f_n(U) \rangle, \label{eq:Gsuma} \\
  f_n(U) &=& \mbox{tr}\, C_1  T'(U R')^{n-1} U T  C_2 T^{\dagger} U^{\dagger} 
      (R'^{\dagger} U^{\dagger})^{n-1} T'^{\dagger}. \label{eq:Gsumb}
\end{eqnarray}
\end{mathletters}%
The average of the polynomial function $f_n(U)$ can be calculated diagrammatically. We represent $f_n(U)$ by the top diagram in Fig.\ \ref{fig:QDWL1}. The average over the matrix $U$ is done as follows.

The leading contribution, which is of order $M$, comes from the diagrams with the largest number of $T$- and $U$-cycles. For a polynomial of the type (\ref{eq:QDWLcondsum}) (all $U$'s are on one side of the $U^{\dagger}$'s), these diagrams have a ``ladder'' structure (see bottom diagram in Fig.\ \ref{fig:QDWL1}). The ladder diagrams contain $n$ $U$-cycles and $n+1$ $T$-cycles. Their weight is $W_{1}^{n} = M^{-n} + {\cal O}(M^{-n-1})$, resulting in
\begin{eqnarray}  \label{eq:fn}
  \langle f_n(U) \rangle &=& M^{-n}\, \mbox{tr}\, T'^{\dagger} C_1 T'\, (\mbox{tr}\, R' R'^{\dagger})^{n-1}\, \mbox{tr}\, T C_2 T^{\dagger} + {\cal O}(1).
\end{eqnarray}
Summation of the series (\ref{eq:Gsum}) yields $\langle G \rangle$ to leading order in $M$,
\begin{eqnarray}
  \langle G/G_0 \rangle &=& {(\mbox{tr}\, T'^{\dagger} C_1 T')(\mbox{tr}\, T C_2 T^{\dagger}) \over
      M - \mbox{tr}\, R' R'^{\dagger}} + {\cal O}(1) \nonumber \\ &=&
    {(N_1 - \mbox{tr}\, \bar S^{\dagger} C_1 \bar S)
      (N_2 - \mbox{tr}\, \bar S C_2 \bar S^{\dagger}) \over
      M - \mbox{tr}\, \bar S \bar S^{\dagger}} + {\cal O}(1). \label{eq:GavgResult}
\end{eqnarray}
In the second equality we have used the unitarity of the matrix $\Sigma$ defined in Eq.\ (\ref{eq:SigmaTdef}). 

The weak-localization correction is the ${\cal O}(1)$ contribution to $\langle G \rangle$. In general, an ${\cal O}(1)$ contribution to the average conductance can have two sources: (i) a higher order contribution to the weight $W_{c_1,\ldots,c_k}$ of the leading-order diagrams, and (ii) higher order diagrams. In the absence of time-reversal symmetry both contributions are absent: (i) $W_{1} = M^{-1}$ has no ${\cal O}(M^{-2})$ term, and (ii) there are no diagrams of order $1$. 

\begin{figure}
\hspace{0.2\hsize}
\epsfxsize=0.6\hsize
\epsffile{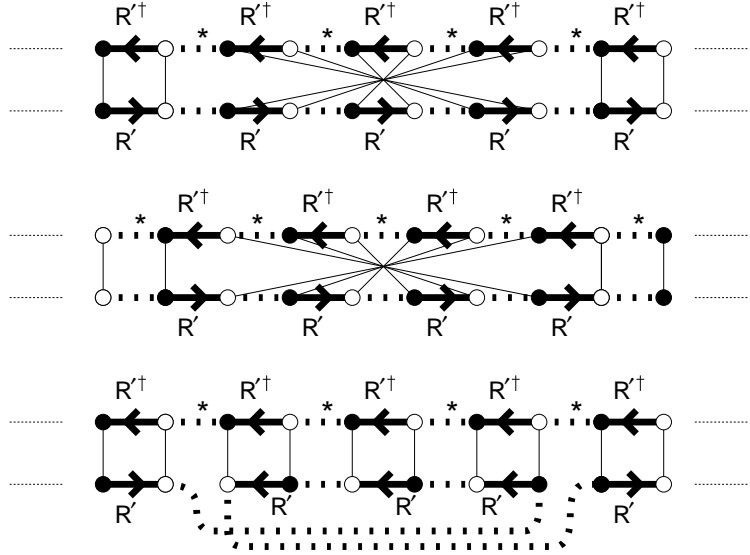}
\bigskip

\caption{\label{fig:QDWL2} Top and middle: Two maximally crossed diagrams contributing to the weak-localization correction to the average conductance. The right and left parts of the diagram have a ladder structure; Bottom: The maximally crossed part of the top diagram redrawn as a ladder diagram.}
\end{figure}%

The situation is different in the presence of time-reversal symmetry. We discuss the case $\beta=1$ in which there is no spin-orbit scattering. The case $\beta=4$ then follows from the translation rule of Sec.\ \ref{sec:trans}. In the presence of time-reversal symmetry, (i) the coefficient $W_{1} = M^{-1} - M^{-2} + \ldots$ has an ${\cal O}(M^{-2})$ term, and (ii) there are diagrams of order $1$. The first contribution is a correction $n M^{-n-1}$ to the weight $M^{-n}$ in Eq.\ (\ref{eq:fn}). Summation over $n$ yields the first correction to Eq.\ (\ref{eq:GavgResult}),
\begin{eqnarray} \label{eq:deltaG1}
  \delta G_1 &=& 
   - {(\mbox{tr}\, T'^{\dagger} C_1 T')(\mbox{tr}\, T C_2 T^{\dagger}) \over
      (M - \mbox{tr}\, R' R'^{\dagger})^2}.
\end{eqnarray}
The second contribution is from diagrams which are obtained from the ladder diagrams by reversing the order of the contractions in a part of the diagram. The central part of the diagram is ``maximally crossed'', the left and right ends are ladders (see Fig.\ \ref{fig:QDWL2}). In disordered systems, the ladder diagrams are known as diffusons, while the maximally crossed diagrams are known as cooperons. The maximally crossed diagrams are not allowed in the absence of time-reversal symmetry, because dots of different color are connected by thin lines (violating rule 2 in Sec.\ \ref{sec:DIAGR}). A maximally crossed diagram can be redrawn as a ladder diagram by flipping one of the horizontal lines along a vertical axis (bottom diagram in Fig.\ \ref{fig:QDWL2}). 

\begin{figure}
\hspace{0.1\hsize}
\epsfxsize=0.8\hsize
\epsffile{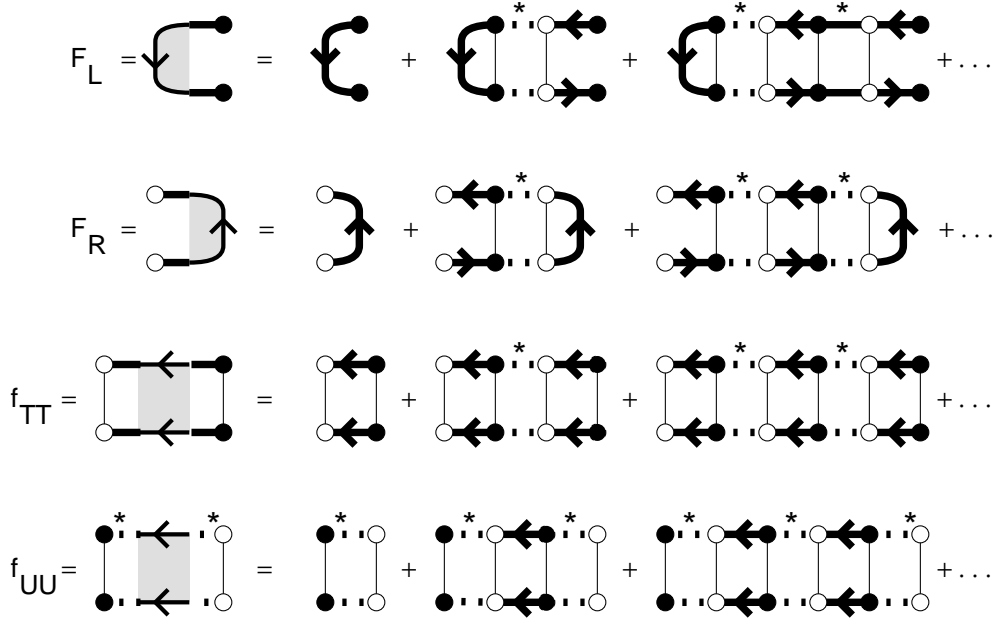}
\medskip

\caption{\label{fig:QDFF1} Diagrammatic representation of Eqs.\ (\protect\ref{eq:FLFRdef}) and (\protect\ref{eq:fTTfUUdef}).}
\end{figure}

\begin{figure}
\hspace{0.0\hsize}
\epsfxsize=0.99\hsize
\epsffile{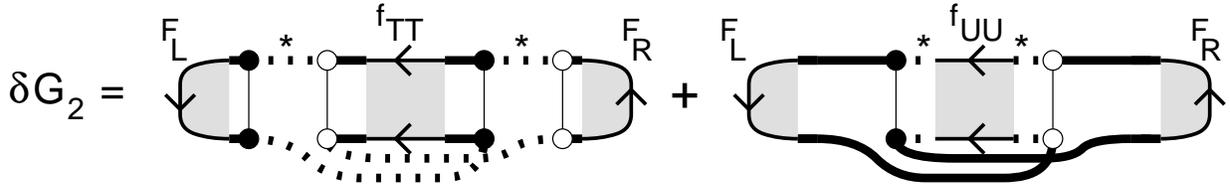}
\bigskip

\caption{\label{fig:QDWL4} Diagrammatic representation of the weak-localization correction $\delta G_2$ from the maximally crossed diagrams. The total correction $\delta G = \delta G_1 + \delta G_2$ contains also a contribution $\delta G_1$ from the weight factors [Eq.\ (\protect\ref{eq:deltaG1})]}
\end{figure}

In the maximally crossed diagrams all cycles but one have minimum length. The cycle with the exceptional length can be a $U$-cycle (top diagram in Fig.\ \ref{fig:QDWL2}), or a $T$-cycle (middle diagram). To evaluate these diagrams, we need to introduce some more notation (see Fig.\ \ref{fig:QDFF1}). We denote the left and right ladder diagrams by matrices $F_{\rm L}$ and $F_{\rm R}$,
\begin{mathletters} \label{eq:FLFRdef}
\begin{eqnarray}
  F_{\rm L} &=& T'^{\dagger} C_1 T' + \sum_{n=1}^{\infty} M^{-n}
    (\mbox{tr}\, T'^{\dagger} C_1 T') (\mbox{tr}\, R'^{\dagger} R')^{n-1}
    R'^{\dagger} R' \nonumber \\ &=&  
    T'^{\dagger} C_1 T' + 
    \left( {\mbox{tr}\, T'^{\dagger} C_1 T' \over M - \mbox{tr}\, R'R'^{\dagger}} \right) 
    R'^{\dagger} R', \\
  F_{\rm R} &=&  T C_2 T^{\dagger} + \sum_{n=1}^{\infty} M^{-n} 
    R' R'^{\dagger} (\mbox{tr}\, R' R'^{\dagger})^{n-1}
    (\mbox{tr}\, T C_2 T^{\dagger}) \nonumber \\ &=&  
    T C_2 T^{\dagger} + R' R'^{\dagger}
    \left({\mbox{tr}\, T C_2 T^{\dagger} \over M - \mbox{tr}\, R'R'^{\dagger}} \right).
\end{eqnarray}
\end{mathletters}%
The scalars $f_{UU}$ and $f_{TT}$ represent the maximally crossed part of the diagram,
\begin{mathletters} \label{eq:fTTfUUdef}
\begin{eqnarray}
  f_{TT} &=& \sum_{n=0}^{\infty} M^{-n} 
    (\mbox{tr}\, R'R'^{\dagger})^{n+1} =  
    {M\, \mbox{tr}\, R R'^{\dagger} \over M - \mbox{tr}\, R'R'^{\dagger}}, \\
  f_{UU} &=& \sum_{n=0}^{\infty} M^{-n-1} 
    (\mbox{tr}\, R'R'^{\dagger})^{n} =  
    {1 \over M - \mbox{tr}\, R'R'^{\dagger}}.
\end{eqnarray}
\end{mathletters}%
We used the symmetry of $R'$ to replace $R'^{\rm T}$ by $R'$. With this notation we may draw the contribution $\delta G_2$ to the weak-localization correction from the maximally crossed diagrams as in Fig.\ \ref{fig:QDWL4}. It evaluates to
\begin{eqnarray} 
  \delta G_2 = -M^{-3}\, \mbox{tr}\, F_{\rm L}^{\vphantom{\rm T}} f_{TT}^{\vphantom{\rm T}}\, \mbox{tr}\, F_{\rm R}^{\vphantom{\rm T}} + 
    \mbox{tr}\, F_{\rm L}^{\vphantom{\rm T}} f_{UU}^{\vphantom{\rm T}} F_{\rm R}^{\rm T}.
\end{eqnarray}

The total weak-localization correction $\delta G = \delta G_1 + \delta G_2$ becomes
\begin{eqnarray}
  \delta G &=& - (\mbox{tr}\, T^{\dagger} T)^{-3} \left[ (\mbox{tr}\, C_2 T^{\dagger} T)^2\, \mbox{tr}\, C_1 (T^{\dagger} T)^2\,+ (\mbox{tr}\, C_1 T^{\dagger} T)^2\, \mbox{tr}\, C_2 (T^{\dagger} T)^2 \right]. \label{eq:dGb1result}
\end{eqnarray}
Since $T^{\dagger} T = 1 - \bar S^{\dagger} \bar S$ has eigenvalues $\Gamma_n$, we may write the final result for the average conductance in the form
\begin{eqnarray} \label{eq:Gtunnel}
  &&  \langle G/G_0 \rangle = {g_1^{\vphantom{2}} g_1' \over g_1\vphantom{'} + g_1'} +
    \left(1-{2 \over \beta} \right) {g_2^{\vphantom{2}} g_1'^2 + g_2' g_1^2 \over (g_1^{\vphantom{2}} + g_1')^3} +
    {\cal O}(M^{-1}), \\
  && g_p^{\vphantom{2}}  = \sum_{n=1}^{N_1} \Gamma_n^p, \ \ 
  g_p' = \sum_{n=1+N_1}^{M} \Gamma_n^p. \label{eq:gkdef}
\end{eqnarray}
(The $\beta=4$ result follows from the translation rule of Sec.\ \ref{sec:trans}.) The first term in Eq.\ (\ref{eq:Gtunnel}) is the series conductance of the two tunnel conductances $G_0 g_1$ and $G_0 g_1'$. The term proportional to $1-2/\beta$ is the weak-localization correction. In the absence of tunnel barriers one has $g_p = N_1$, $g_p' = N_2$, and the large-$M$ limit of Eq.\ (\ref{eq:QDavgC}) is recovered. In the case of two identical tunnel barriers ($N_1 = N_2 = M/2 \equiv N$, $\Gamma_n = \Gamma_{n+N}$ for $j=1,\ldots,N$), Eq.\ (\ref{eq:Gtunnel}) simplifies to
\begin{eqnarray} \label{eq:GtunnelId}
  \langle G/G_0 \rangle &=& \frac{1}{2} g_1 +
    \left(1-{2 \over \beta} \right) {g_2 \over 4\, g_1} +
    {\cal O}(M^{-1}).
\end{eqnarray}
Eq.\ (\ref{eq:GtunnelId}) was previously obtained by Iida, Weidenm\"uller and Zuk \cite{IWZ}.
If all $\Gamma_n$'s are equal to $\Gamma$, Eq.\ (\ref{eq:GtunnelId}) simplifies further to $\langle G/G_0 \rangle = \case{1}{2} N \Gamma + \case{1}{4} (1 - 2/\beta) \Gamma$.

\subsection{Conductance fluctuations} \label{sec:fluctQD}

\begin{figure}
\hspace{0.1\hsize}
\epsfxsize=0.8\hsize
\epsffile{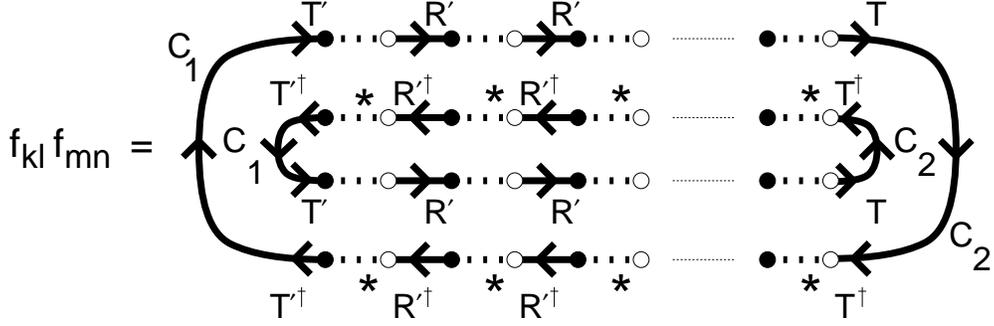}
\medskip

\caption{\label{fig:QDCF1} Diagrammatic representation of a term contributing to $G^2$, and hence to the variance (\protect\ref{eq:VarG}) of the conductance.}
\end{figure}

We seek the effect of tunnel barriers on the variance of the conductance, $\mbox{var}\, G = \langle G^2 \rangle - \langle G \rangle^2$. We consider $\beta=1$ and $2$ first, and translate to $\beta=4$ in the end. Using the decomposition (\ref{eq:Sparam}) we write the variance in the form 
\begin{mathletters} \label{eq:VarG}
\begin{eqnarray} 
 && \mbox{var}\, G/G_0 =
     \mbox{var}\, (\mbox{tr}\, C_1 \delta S C_2 \delta S^{\dagger}) =
  \sum_{k,l,m,n \ge 1} \mbox{covar}\, (f_{kl}, f_{mn}), \\
 && f_{kl} = \mbox{tr}\, C_{1} T' (U R')^{k-1} U T C_{2} T^{\dagger} U^{\dagger} (R'^{\dagger} U^{\dagger})^{l-1} T'^{\dagger} \label{eq:Gfluct}.
\end{eqnarray}  
\end{mathletters}%
Since the number $U$'s and $U^{*}$'s must be equal for a non-zero average, $\mbox{covar}\, (f_{kl},f_{mn}) \equiv \langle f_{kl} f_{mn} \rangle - \langle f_{kl} \rangle \langle f_{mn} \rangle = 0$ unless $k+m=l+n$. 
Diagrammatically, we represent $f_{kl} f_{mn}$ by Fig.\ \ref{fig:QDCF1}. The diagram consists of an inner loop, corresponding to $f_{kl}$, and an outer loop, corresponding to $f_{mn}$. The covariance of $f_{kl}$ and $f_{mn}$ is given by the connected diagrams. We call a diagram ``connected'' if (i) the partition of the $U$-cycles contains a group which consists of $U$-cycles from the inner and the outer part, or (ii) the diagram contains a cycle (a $U$-cycle or a $T$-cycle) connecting the inner and outer loops.

We first compute the contribution from diagrams which are connected only because of (i), i.e.\ diagrams in which all $U$-cycles and $T$-cycles belong either to the inner or outer loop. The contribution from such a diagram is maximal, if the $U$-cycles are partitioned in groups which are as small as possible. The optimal partition consists of groups of size $1$, except for a single group of size $2$, which contains one $U$-cycle from the inner and one from the outer loop. Furthermore, the total number of cycles is maximal if both the inner and outer loops are ladder diagrams. This requires $k=l$ and $m=n$. The covariance from this diagram is 
\begin{eqnarray}
  \mbox{covariance} &=&
    k m \, \delta_{k l} \delta_{m n} W_{1,1}^{\vphantom{2}} W_{1}^{k+m-2}\,
 \nonumber \\ && \mbox{} \times
  (\mbox{tr}\, T'^{\dagger} C_{1} T')^2\, (\mbox{tr}\, R' R'^{\dagger})^{k+m-2}
    (\mbox{tr}\, T C_{2} T^{\dagger})^2  + {\cal O}(M^{-1}).
\end{eqnarray}
Summing over $k$ and $m$ we obtain the first contribution to $\mbox{var}\, G/G_0$,
\begin{equation} \label{eq:VarG1beta2}
  \mbox{variance}\, = M^{-4}\, (\mbox{tr}\, F_{\rm L}\, \mbox{tr}\, F_{\rm R})^2.
\end{equation}

\begin{table}
\begin{center}
\begin{tabular}{c|cc}
 diagram &  $\beta=1,2$ & $\beta=1$ \\ \hline
 a   &\ $W_{2}^{2}\, (\mbox{tr}\, F_{\rm L})^2\, f_{TT}^2\,
            (\mbox{tr}\, F_{\rm R})^2$ 
     &\ $W_{2}^{2}\, \mbox{tr}\, F_{\rm R}\, \mbox{tr}\, R_{\rm L}\, f_{TT}^2\,
            \mbox{tr}\, F_{\rm L}\, \mbox{tr}\, F_{\rm R}$ \\
 b   &\ $W_{3}\, (\mbox{tr}\, F_{\rm L})^2\, f_{TT}\,
            (\mbox{tr}\, F_{\rm R})^2$
     &\ $W_{3}\, \mbox{tr}\, F_{\rm R}\, \mbox{tr} F_{\rm L}\, f_{TT}\,
            \mbox{tr}\, F_{\rm L}\, \mbox{tr}\, F_{\rm R}$ \\ 
 c   &\ $W_{2}\, (\mbox{tr}\, F_{\rm L})^2\, f_{TU}^2\, 
            \mbox{tr}\, F_{\rm R}^2$
     &\ $W_{2}\, \mbox{tr}\, F_{\rm R}\, \mbox{tr}\, F_{\rm L}\, f_{TU}^2\,
            \mbox{tr}\, F_{\rm L}^{\rm T} F_{\rm R}\,$ \\
 d   &\ $\mbox{tr}\, H H^{\dagger}\, f_{UU}$
     &\ $\mbox{tr}\, H^{\rm T} H^{\dagger}\, f_{UU}$ \\ \hline
 e   &\ $W_{2}\, \mbox{tr}\, F_{\rm L}^2\, f_{UT}^2\, 
            (\mbox{tr}\, F_{\rm R})^2$ 
     &\ $W_{2}\, \mbox{tr}\, F_{\rm R}^{\rm T} F_{\rm L} f_{UT}^2\, 
            \mbox{tr}\, F_{\rm L}\, \mbox{tr}\, F_{\rm R}$ \\
 f   &\ $W_{3}\, (\mbox{tr}\, F_{\rm L})^2\, f_{TT}\,
            (\mbox{tr}\, F_{\rm R})^2$
     &\ $W_{3}\, \mbox{tr}\, F_{\rm R}\, \mbox{tr} F_{\rm L}\, f_{TT}\,
            \mbox{tr}\, F_{\rm L}\, \mbox{tr}\, F_{\rm R}$ \\
 g   &\ $\mbox{tr}\, F_{\rm L}^2\, f_{UU}^2\,
            \mbox{tr}\, F_{\rm R}^2$
     &\ $\mbox{tr}\, F_{\rm R}^{\rm T} F_{\rm L}\, f_{UU}^2\,
            \mbox{tr}\, F_{\rm L}^{\rm T} F_{\rm R}$ \\ 
 h   &\ $\mbox{tr}\, H^{\dagger} H\, f_{UU}$
     &\ $\mbox{tr}\, H^{*} H\, f_{UU}$ \\ \hline
 i   &\ $W_{2}^{2}\, \mbox{tr}\, F_{\rm L}\, \mbox{tr}\, F_{\rm R}\, f_{TT}^2\,
            \mbox{tr}\, F_{\rm L}\, \mbox{tr}\, F_{\rm R}$ 
     &\ $W_{2}^{2}\, \mbox{tr}\, F_{\rm R}\, \mbox{tr}\, F_{\rm L}\, f_{TT}^2\,
            \mbox{tr}\, F_{\rm L}\, \mbox{tr}\, F_{\rm R}$ \\
 j   &\ $W_{2}^{2}\, \mbox{tr}\, F_{\rm L}\, \mbox{tr}\, F_{\rm R}\, f_{TT}^2\,
            \mbox{tr}\, F_{\rm L}\, \mbox{tr}\, F_{\rm R}$ 
     &\ $W_{2}^{2}\, \mbox{tr}\, F_{\rm L}\, \mbox{tr}\, F_{\rm R}\, f_{TT}^2\,
            \mbox{tr}\, F_{\rm R}\, \mbox{tr}\, F_{\rm L}$ \\
 k   &\ $W_{2}\, \mbox{tr}\, H R'^{\dagger}\, f_{TU} f_{UT}\,
            \mbox{tr}\, F_{\rm L}\, \mbox{tr}\, F_{\rm R}$
     &\ $W_{2}\, \mbox{tr}\, H^{\rm T} R'^{\dagger}\, f_{TU} f_{UT}\,
            \mbox{tr}\, F_{\rm L}\, \mbox{tr}\, F_{\rm R}$ \\
 l   &\ $W_{2}\, \mbox{tr}\, F_{\rm L}\, \mbox{tr}\, F_{\rm R}\, f_{TU} f_{UT}\,
            \mbox{tr}\, H^{\dagger} R'$
     &\ $W_{2}\, \mbox{tr}\, F_{\rm L}\, \mbox{tr}\, F_{\rm R}\, f_{TU} f_{UT}\,
            \mbox{tr}\, H^{*} R'$ \\ \hline
 m   &\ $W_{2}\, \mbox{tr}\, F_{\rm L}\, \mbox{tr}\, F_{\rm R}\, f_{TU} 
            f_{UT}\, \mbox{tr}\, R' H^{\dagger}$
     &\ $W_{2}\, \mbox{tr}\, F_{\rm R}\, \mbox{tr}\, F_{\rm L}\, f_{TU}
            f_{UT}\, \mbox{tr}\, R'^{\rm T} H^{\dagger}$ \\
 n   &\ $W_{2}\, \mbox{tr}\, R'^{\dagger} H\, f_{TU} f_{UT}\,
            \mbox{tr}\, F_{\rm L}\, \mbox{tr}\, F_{\rm R}$
     &\ $W_{2}\, \mbox{tr}\, R'^{*} H\, f_{TU} f_{UT}\,
            \mbox{tr}\, F_{\rm L}\, \mbox{tr}\, F_{\rm R}$ \\
 o   &\ $\mbox{tr}\, H R'^{\dagger}\, f_{UU}^2\, \mbox{tr}\, R' H^{\dagger}$ 
     &\ $\mbox{tr}\, H^{\rm T} R'^{\dagger}\, f_{UU}^2\,
          \mbox{tr}\, R'^{\rm T} H^{\dagger}\, $ \\
 p   &\ $\mbox{tr}\, R'^{\dagger} H\, f_{UU}^2 \mbox{tr}\, H^{\dagger} R'$
     &\ $\mbox{tr}\, R'^{*} H f_{UU}^2\, \mbox{tr}\, H^{*} R'$ \\
\end{tabular}
\end{center}

\caption{\label{tab:1} Contribution to $\mbox{var}\, G/G_0$ from the connected diagrams of Fig.\ \protect\ref{fig:QDCF7}.}
\end{table}

\begin{figure}
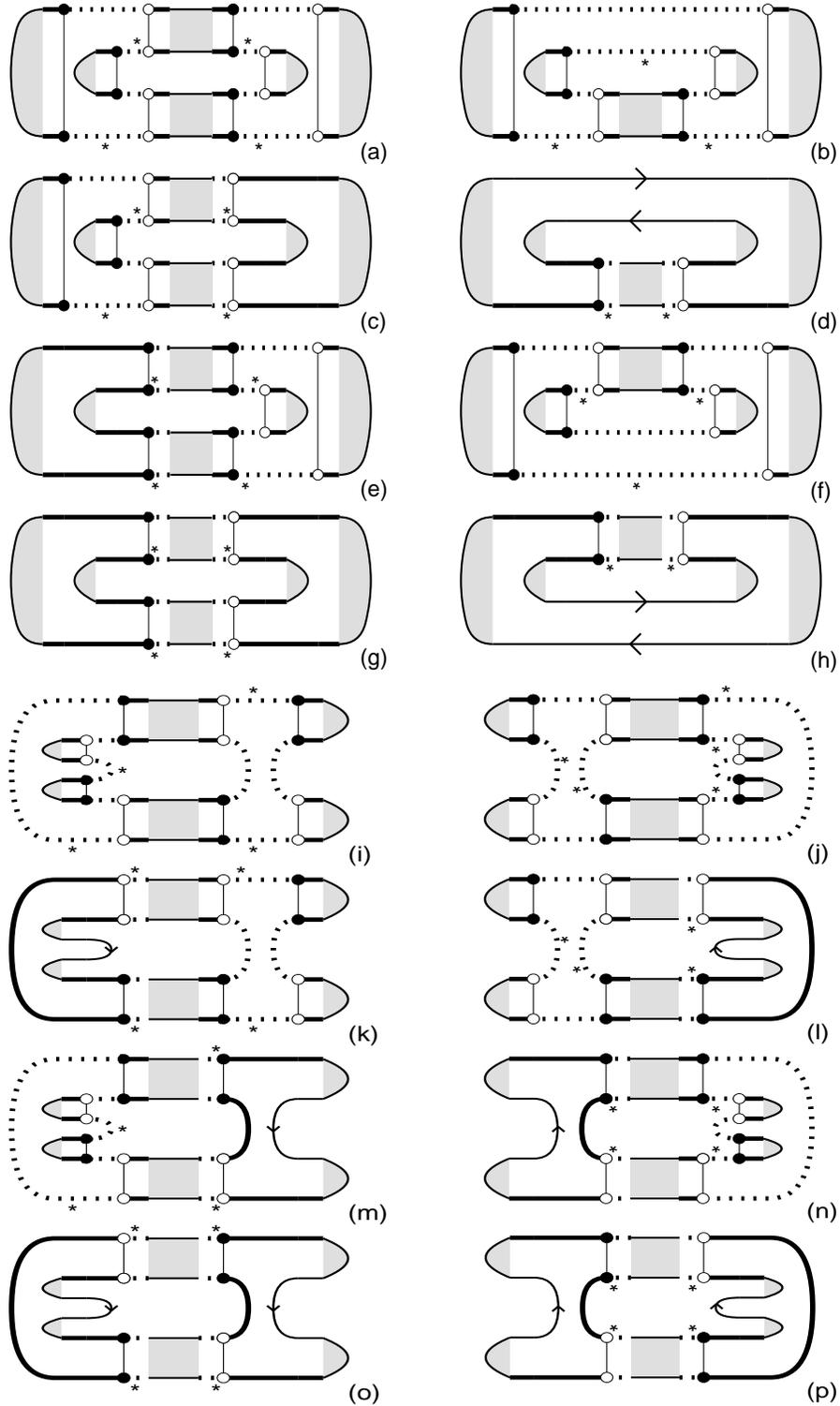

\hspace{0.12\hsize}
\epsfxsize=0.33\hsize
\epsffile{afigQDCF4.eps}
\hspace{0.05\hsize}
\epsfxsize=0.33\hsize
\epsffile{afigQDCF7.eps}

\hspace{0.12\hsize}
\epsfysize=0.63\hsize
\epsfxsize=0.33\hsize
\epsffile{afigQDCF5.eps}
\hspace{0.05\hsize}
\epsfysize=0.63\hsize
\epsfxsize=0.33\hsize
\epsffile{afigQDCF6.eps}
\medskip

\caption{\label{fig:QDCF7} The 16 connected diagrams which contribute to the variance of the conductance. The shaded parts are defined in Figs.\ \protect\ref{fig:QDFF1} and \protect\ref{fig:QDFF2}. These diagrams contribute for $\beta=1$ and $2$. For $\beta=1$ there are 16 more diagrams, obtained by flipping the inner loop around a vertical axis (diagram a--h) or around a horizontal axis (i--p), so that ladders become maximally crossed.}
\end{figure}%

\begin{figure}
\hspace{0.1\hsize}
\epsfxsize=0.8\hsize
\epsffile{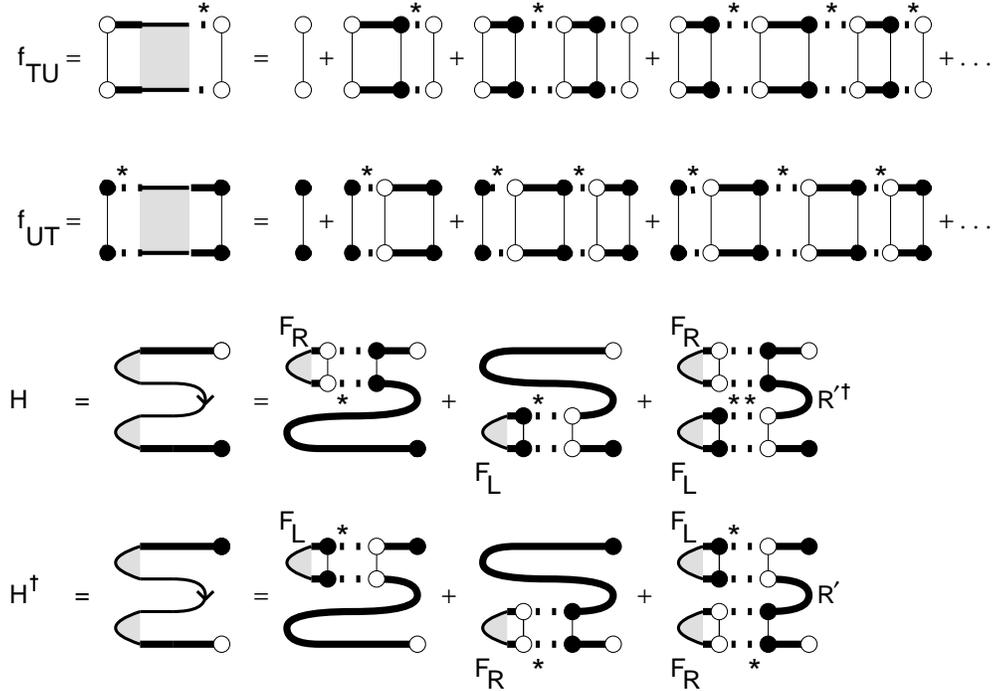}
\medskip

\caption{\label{fig:QDFF2} Diagrammatic representation of Eq.\ (\protect\ref{eq:fUTfTUfLRfRLdef}).}
\end{figure}%

The second contribution, consisting of diagrams in which the inner and outer loops are connected by $T$- or $U$-cycles, is of maximal order if the partition of the $U$-cycles involves only groups of size $1$. For $\beta=2$ there are 16 connected diagrams of maximal order. They are shown in Fig.\ \ref{fig:QDCF7}, and their contribution to $\mbox{var}\, G/G_0$ is tabulated in Table \ref{tab:1}. The shaded areas indicate ladder parts of the diagram (see Figs.\ \ref{fig:QDFF1} and \ref{fig:QDFF2}). The matrices $F_{\rm L}$ and $F_{\rm R}$, and the scalars $f_{UU}$ and $f_{TT}$ are defined in Eqs.\ (\ref{eq:FLFRdef}) and (\ref{eq:fTTfUUdef}). The definitions of the matrix $H$ and of the scalars $f_{UT}$ and $f_{TU}$ are
\begin{mathletters} \label{eq:fUTfTUfLRfRLdef}
\begin{eqnarray} 
 && f_{UT} = f_{TU} = \sum_{n=1}^{\infty} M^{-n} 
    (\mbox{tr}\, R'R'^{\dagger})^{n} =  
    {\mbox{tr}\, R'R'^{\dagger} \over M - \mbox{tr}\, R'R'^{\dagger}}, \\
 && H        M^{-1} (\mbox{tr}\, F_{\rm R}) R' T'^{\dagger} C_{1} T' +
             M^{-1} (\mbox{tr}\, F_{\rm L}) T C_{2} T^{\dagger} R' +
             M^{-2} (\mbox{tr}\, F_{\rm L}) (\mbox{tr}\, F_{\rm R}) R' R'^{\dagger} R'.
\end{eqnarray}
\end{mathletters}%
In the presence of time-reversal symmetry ($\beta=1$), the matrix $U$ is symmetric. Diagrammatically, this means that no distinction is made between black and white dots. In addition to the 16 diffuson-like diagrams of Fig.\ \ref{fig:QDCF7}, 16 more cooperon-like diagrams contribute. These are obtained from the diagrams of Fig.\ \ref{fig:QDCF7} by flipping the inner loop around a vertical (Fig.\ \ref{fig:QDCF7}a--h) or horizontal (Fig.\ \ref{fig:QDCF7}i--p) axis, so that segments with a ladder structure become maximally crossed. Their contributions are listed in Table \ref{tab:1}. The contributions from the individual diffuson-like and cooperon-like diagrams are different. The total contribution to $\mbox{var}\, G$ from diffuson-like and cooperon-like diagrams is the same.

The final result for the variance of $G$ is 
\begin{eqnarray}
  \mbox{var}\, G/G_0 &=& 2 \beta^{-1}
     \left(g_1^{\vphantom{2}} + g_1'\right)^{-6}
     \left(2 g_1^4 g_1'^2 + 4 g_1^3 g_1'^3 - 
     4 g_1^2 g_2^{\vphantom{2}} g_1'^3 +
     2 g_1^2 g_1'^4 -
     2 g_1^{\vphantom{2}} g_2^{\vphantom{2}} g_1'^4 
       \right. \nonumber \\ && \left. \mbox{} +
     3 g_2^2 g_1'^4 - 
     2 g_1^{\vphantom{2}} g_3^{\vphantom{2}} g_1'^4 + 
     2 g_2^{\vphantom{2}} g_1'^5 -
     2 g_3^{\vphantom{2}} g_1'^5 + 2 g_1^5 g_2' -
     2 g_1^4 g_1' g_2' 
       \right. \nonumber \\ && \left. \mbox{} -
     4 g_1^3 g_1'^2 g_2' + 
     6 g_1^2 g_2^{\vphantom{2}} g_1'^2 g_2' + 3 g_1^4 g_2'^2 - 
     2 g_1^5 g_3' - 2 g_1^4 g_1' g_3' \right). \label{eq:QDVarDiagr}
\end{eqnarray}
One verifies that the large-$N$ limit of Eq.\ (\ref{eq:QDvarC}) is recovered in the absence of tunnel barriers. For the special case of identical tunnel barriers ($g_p = g_p')$, this simplifies to
\begin{eqnarray}
  \mbox{var}\, G/G_0 &=& (8 \beta\, g_1^2)^{-1} \left( {2 g_1^2 - 2 g_1^{\vphantom{2}} g_2^{\vphantom{2}} + 3 g_2^2 - 2 g_1^{\vphantom{2}} g_3^{\vphantom{2}} } \right),
\end{eqnarray}
in agreement with Ref.\ \onlinecite{IWZ}.
If all transmission eigenvalues $\Gamma_n \equiv \Gamma$ are equal, one has $\mbox{var}\, G/G_0 = (8 \beta)^{-1} [{1 + (1 - \Gamma)^2}]$. A high tunnel barrier ($\Gamma \ll 1$) thus doubles the variance.

\subsection{Density of transmission eigenvalues} \label{sec:denstrans}

The transmission eigenvalues $T_n \in [0,1]$ are the $N_1$ eigenvalues of the matrix product $s_{12}^{\vphantom{\dagger}} s_{12}^{\dagger}$. Without loss of generality we may assume that $N_1 \le N_2$. The matrix product $s_{21}^{\vphantom{\dagger}} s_{21}^{\dagger}$ then has the same $N_1$ eigenvalues as $s_{12}^{\vphantom{\dagger}} s_{12}^{\dagger}$, plus $N_2 - N_1$ eigenvalues equal to zero. The $N_1$ non-zero transmission eigenvalues appear as the diagonal elements of the diagonal matrix $T$ in the polar decomposition of the scattering matrix
\begin{equation}
  S = \left( \begin{array}{cc} s_{11} & s_{12} \\ s_{21} & s_{22} \end{array} \right) =
 \left( \begin{array}{cc} v & 0 \\ 0 & w \end{array} \right)
      \left( \begin{array}{ccc} \sqrt{1-T} & 0 & i \sqrt{T} \\
      0 & \openone & 0 \\
        i \sqrt{T} & 0 & \sqrt{1-T} \end{array} \right)
      \left( \begin{array}{cc} v' & 0 \\ 
        0 & w' \end{array} \right). \label{eq:polardecomp12}
\end{equation}
Here $v$ and $v'$ ($w$ and $w'$) are $N_1 \times N_1$ ($N_2 \times N_2$) unitary matrices and $\openone$ is the $N_2 - N_1$ dimensional unit matrix. If $N_1 = N_2$, Eq.\ (\ref{eq:polardecomp12}) simplifies to Eq.\ (\ref{eq:polardecomp1}).

Sofar we have only studied the conductance $G = G_0 \sum_{n} T_n$. The leading contribution to the average conductance comes from ladder diagrams. If we wish to average transport properties of the form $A = \sum_n a(T_n)$ (so-called linear statistics on the transmission eigenvalues), we need to know the density $\rho(T)$ of the transmission eigenvalues $T_n$. The leading-order contribution to the transmission-eigenvalue density is given by a larger class of diagrams, as we now discuss.

The density $\rho(T) = \langle \sum_{n=1}^{N_1} \delta(T-T_n) \rangle$ of the transmission eigenvalues follows from the matrix Green function $F(z)$:
\begin{mathletters}
\begin{eqnarray} \label{eq:Fdef}
  F(z) &=& \left\langle C_1 (z - S C_2 S^{\dagger} C_1)^{-1} \right\rangle, \\
  \rho(T) &=& -\pi^{-1} \mbox{Im}\, \mbox{tr}\, F(T + i \epsilon),
\end{eqnarray}
\end{mathletters}%
where $\epsilon$ is a positive infinitesimal. We first compute $\rho(T)$ in the absence of tunnel barriers, when the result is known from other methods\cite{BeenakkerReview,BarangerMello,JPB,NazarovR}. Then we include the tunnel barriers, when the result is not known.

\begin{figure}
\hspace{0.1\hsize}
\epsfxsize=0.8\hsize
\epsffile{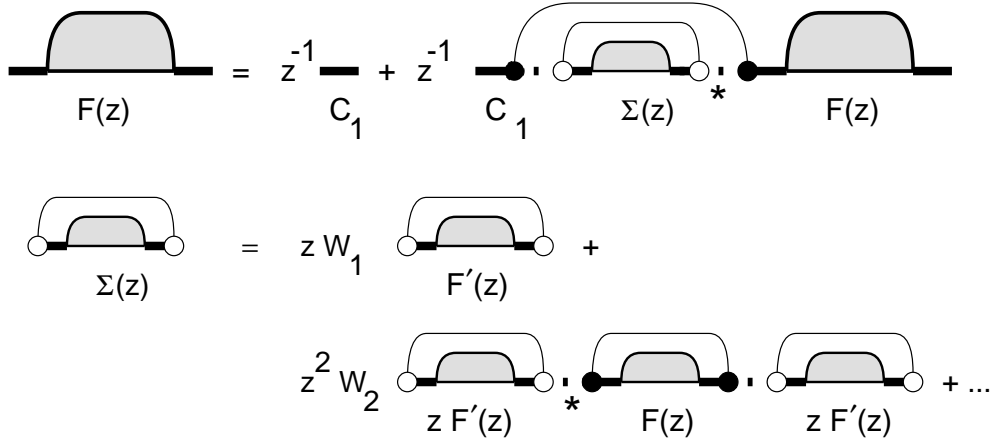}
\medskip

\caption{\label{fig:QDDT3}Diagrammatic representation of the Dyson equation (\protect\ref{eq:FzSelfEq}) for $F(z)$.}
\end{figure}

In the absence of tunnel barriers, the scattering matrix $S$ is distributed according to the circular ensemble, so that averaging amounts to integrating over the unitary group. We compute $F(z)$ as an expansion in powers of $1/z$,
\begin{equation}
  F(z) = \sum_{n=0}^{\infty} 
    \langle  C_1 z^{-1} (S C_2 S^{\dagger} C_1 z^{-1})^{n} \rangle.
\end{equation}
We will also need the Green function
\begin{eqnarray}
  F'(z) &=& \langle C_2 (z - S^{\dagger} C_1 S C_2)^{-1} \rangle 
  = \sum_{n=0}^{\infty} 
    \langle  C_2 z^{-1} (S^{\dagger} C_1 S C_2 z^{-1})^{n} \rangle.
\end{eqnarray}
\begin{figure}
\hspace{0.2\hsize}
\epsfxsize=0.6\hsize
\epsffile{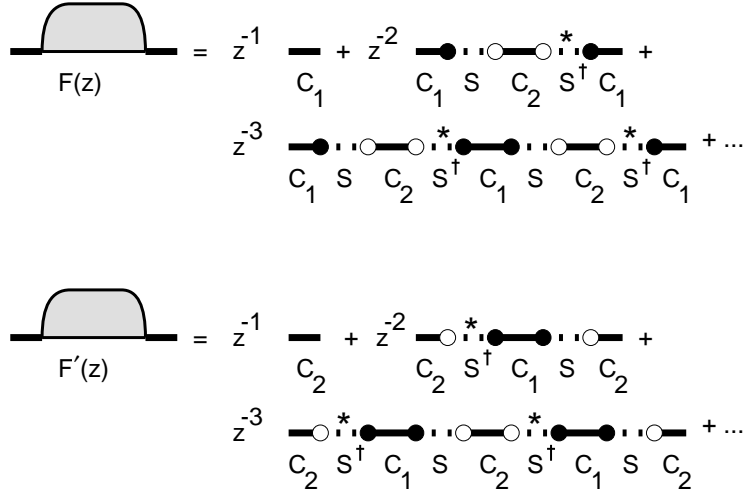}
\medskip

\caption{\label{fig:QDDT2} Diagrammatic representation of the Green functions for the density of transmission eigenvalues.}
\end{figure}%
The two Green functions $F$ and $F'$ are represented diagrammatically in Fig.\ \ref{fig:QDDT2}. A diagram contributes to leading order [which is ${\cal O}(1)$] if the number of $T$-and $U$-cycles is maximal. That is the case if the diagram is {\em planar}, meaning that the thin lines do not cross. The ladder diagrams are a subset of the planar diagrams. Planar diagrams have been studied in the context of the diagrammatic evaluation of integrals over Hermitian matrices, in particular for the Gaussian ensemble \cite{Pandey,BrezinZee}. For the Gaussian ensemble, only planar diagrams with $U$-cycles of unit length have to be taken into account. Summation over all these diagrams results in a self-consistency or Dyson equation for $F(z)$, which solves the problem \cite{BrezinZee}. For an integral of unitary matrices, $U$-cycles of arbitrary length need to be taken into account, as is shown diagrammatically in Fig.\ \ref{fig:QDDT3}. The corresponding Dyson equation is
\begin{mathletters} \label{eq:FzSelfEq}
\begin{eqnarray} 
  \lefteqn{F (z)  =  z^{-1} C_1 + z^{-1} C_1 \Sigma (z) F (z),}
  \hphantom{F'(z)  =  z^{-1} C_2 + z^{-1} C_2 \Sigma'(z) F'(z),}
  \label{eq:FzSelfEqa} \ && \
  \Sigma (z) =
    \sum_{n=1}^{\infty} W_{n} \left[z\, \mbox{tr}\, F'(z)\right]^{n} [\mbox{tr}\, F (z)]^{n-1},
    \label{eq:FzSelfEqc} \\
  F'(z)  =  z^{-1} C_2 + z^{-1} C_2 \Sigma'(z) F'(z) \label{eq:FzSelfEqb},
\ && \
  \Sigma'(z) =
    \sum_{n=1}^{\infty} W_{n} [z\, \mbox{tr}\, F (z)]^{n} [\mbox{tr}\, F'(z)]^{n-1} .
    \label{eq:FzSelfEqd} 
\end{eqnarray}
\end{mathletters}%
In terms of the generating function
\begin{eqnarray} \label{eq:hdef}
  h(z) &=& \sum_{n=1}^{\infty} W_{n} z^{n-1} = {1 \over 2 z} \left( \sqrt{M^2 + 4 z} - M \right),
\end{eqnarray}
we may rewrite Eq.\ (\ref{eq:FzSelfEq}) as
\begin{mathletters} \label{eq:FzSelfEq2}
\begin{eqnarray}  
  F (z) = C_1 ( z - \Sigma(z) C_1)^{-1},\ && \ \Sigma (z) = h \biglb( z\, \mbox{tr}\, F (z)\, \mbox{tr}\, F'(z) \bigrb)\, z\, \mbox{tr}\, F'(z), \\
  F'(z) = C_2 ( z - \Sigma(z) C_2)^{-1},\ && \ \Sigma'(z) = h \biglb( z\, \mbox{tr}\, F (z)\, \mbox{tr}\, F'(z) \bigrb)\, z\, \mbox{tr}\, F (z).
\end{eqnarray}
\end{mathletters}%

In the derivation of Eq.\ (\ref{eq:FzSelfEq2}) we did not use the particular form of the matrices $C_1$ and $C_2$. As a check we may choose $C_1 = C_2 = 1$, so that $F(z) = F'(z) = (z-1)^{-1}$, and verify that Eq.\ (\ref{eq:FzSelfEq2}) holds.

The solution of Eq.\ (\ref{eq:FzSelfEq2}) is
\begin{mathletters}
\begin{eqnarray}
  \mbox{tr}\, F(z)  &=& {N_1 - N_2 \over 2 z} + {\sqrt{M^2 z - (N_1 - N_2)^2} \over 2 z \sqrt{z-1}}, \\
  \mbox{tr}\, F'(z) &=& {N_2 - N_1 \over 2 z} + {\sqrt{M^2 z - (N_2 - N_1)^2} \over 2 z \sqrt{z-1}}.
\end{eqnarray}
\end{mathletters}%
The resulting density of transmission eigenvalues is
\begin{equation}
  \rho(T) = {M\sqrt{T - T_{\rm min}} \over 2 \pi T \sqrt{1 - T}} \theta(T - T_{\rm min}), \ \  T_{\rm min} = {(N_1-N_2)^2 \over M^2}, \label{eq:rho}
\end{equation}
in agreement with Refs.\ \onlinecite{BarangerMello,JPB,NazarovR}. (The function $\theta(x)=1$ if $x>0$ and $0$ if $x<0$.)

The weak-localization correction to $\rho(T)$ follows from the ${\cal O}(M^{-1})$ term in the large-$M$ expansion of $F(z)$. As in Sec.\ \ref{sec:QDavg}, it has two contributions: $\delta F_1(z)$, which is due to the sub-leading order term in the large-$M$ expansion of $W_{n}$, and $\delta F_2(z)$, which is due to diagrams of order ${\cal O}(M^{-1})$. In the absence of time-reversal symmetry, both contributions are absent. In the presence of time-reversal symmetry, the sub-leading order term $\delta W_{n} = -M^{-2n} (-4)^{n-1}$ in the large-$M$ expansion of $W_{n}$ [cf.\ Eq.\ (\ref{eq:VNlargeb1})] yields a sub-leading order contribution $\delta h$ to the generating function $h$,
\begin{eqnarray} \label{eq:dhdef}
  \delta h(z) &=& \sum_{n=1}^{\infty} \delta W_{n} z^{n-1}
    = - (M^2 + 4 z)^{-1},
\end{eqnarray}
from which we obtain
\begin{eqnarray}
  \mbox{tr}\, \delta F_1(z) = \case{1}{4}(z-T_{\rm min})^{-1} - \case{1}{4}(z-1)^{-1}.
\end{eqnarray}
The contribution $\delta F_2(z)$ comes from diagrams in which thin lines connect black and white dots. Each such diagram contains the product $C_1 C_2$, which vanishes. Hence, the ${\cal O}(M^{-1})$ contribution to $F(z)$ consists of $\delta F_1(z)$ only. The resulting weak-localization correction to the transmission eigenvalue density is 
\begin{equation} \label{eq:drho}
  \delta \rho(T) = {2 - \beta \over 4\beta} \left[ \delta(T - T_{\rm min} - \epsilon) - \delta(T - 1 + \epsilon) \right],
\end{equation}
in agreement with Refs.\ \onlinecite{BeenakkerReview,JPB}.

We now include tunnel barriers in the leads. Motivated by Nazarov's calculation of the density of transmission eigenvalues in a disordered metal \cite{Nazarov}, we introduce the $2M \times 2M$ matrices
\begin{mathletters}
\begin{eqnarray}
  {\bf S} &=& \left( \begin{array}{cc} S & 0\\ 0 & S^{\dagger} \end{array} \right), \ \ {\bf C} = \left( \begin{array}{cc} 0 & C_2 \\ C_1 & 0 \end{array} \right), \ \ {\bf F}(z) = \left( \begin{array}{cc} 0 & F'(z) \\ F(z) & 0 \end{array} \right), \\
  {\bf T} &=& \left( \begin{array}{cc} T & 0\\ 0 & T^{\dagger} \end{array} \right), \ \ {\bf T'} = \left( \begin{array}{cc} T' & 0\\ 0 & T'^{\dagger} \end{array} \right), \ \ {\bf R'} = \left( \begin{array}{cc} R' & 0\\ 0 & R'^{\dagger} \end{array} \right).
\end{eqnarray}
\end{mathletters}%
Analogous to Eq.\ (\ref{eq:Sparam}), we decompose ${\bf S} = {\bf \bar S} + {\bf \delta S}$, where ${\bf \bar S} = \langle {\bf S} \rangle$ and
\begin{eqnarray}
  {\bf \delta S} = {\bf T}' (1 - {\bf U} {\bf R}')^{-1} {\bf U} {\bf T},
  \ \
 {\bf U} = \left( \begin{array}{cc} U & 0 \\ 0 & U^{\dagger} \end{array} \right)
\end{eqnarray}
is given in terms of a matrix $U$ which is distributed according to the circular ensemble. Because $\bar S$, $C_1$, and $C_2$ commute and $C_1 C_2 = 0$, we may replace $S$ by $\delta S$ in the expression (\ref{eq:Fdef}) for $F(z)$. The result for the matrix Green function ${\bf F}(z)$ is
\begin{eqnarray}
  {\bf F}(z) &=& 
    (2 z)^{-1} \sum_{\pm}
    \left\langle {\bf C} \pm {\bf C} {\bf T}'
    \left[1 - {\bf U}({\bf R}' \pm {\bf T}{\bf C}{\bf T}' z^{-1/2})
    \right]^{-1} {\bf U} {\bf T} {\bf C} z^{-1/2} \right\rangle  \nonumber \\
  &=&
    (2z)^{-1} \sum_{\pm}
    \left[ {\bf C} \pm {\bf A}_{\pm}
     ({\bf F}_{\pm} - {\bf X}_{\pm}) {\bf B}_{\pm} 
    \right]. \label{eq:Fcalres1}
\end{eqnarray}
In the second equation we abbreviated ${\bf X}_{\pm} = {\bf R}' \pm {\bf T}{\bf C}{\bf T}' z^{-1/2}$, ${\bf F}_{\pm} = \langle {\bf X}_{\pm} (1 - {\bf U} {\bf X}_{\pm})^{-1} \rangle$, and defined ${\bf A}_{\pm}$ and ${\bf B}_{\pm}$ such that $  {\bf A}_{\pm} {\bf X}_{\pm} = {\bf C} {\bf T}'$, $  {\bf X}_{\pm} {\bf B}_{\pm} = {\bf T} {\bf C} z^{-1/2}$.

After these algebraic manipulations we are ready to compute ${\bf F}_{\pm}$ by expanding in planar diagrams. The result is a Dyson equation similar to Eq.\ (\ref{eq:FzSelfEq}),
\begin{eqnarray} \label{eq:calfdef}
  {\bf F}_{\pm} =
    {\bf X}_{\pm}\left( 1 + {\bf \Sigma}_{\pm} {\bf F}_{\pm} \right), \ \
  {\bf \Sigma}_{\pm} =
    \sum_{n=1}^{\infty} W_n \left( {\cal P} {\bf F}_{\pm} \right)^{2n - 1},
\end{eqnarray}
where the projection operator ${\cal P}$ acts on a $2M \times 2M$ matrix ${\bf A}$ as
\begin{equation}
  {\bf A} = \left( \begin{array}{cc}
    A_{11} & A_{12} \\ A_{21} & A_{22} \end{array}\right), \ \
  {\cal P} {\bf A} = \left( \begin{array}{cc}
    0 & \openone_M\, \mbox{tr}\, A_{12} \\ \openone_M\, \mbox{tr}\, A_{21} & 0
  \end{array} \right),
\end{equation}
$\openone_M$ being the $M \times M$ unit matrix. The presence of the projection operator ${\cal P}$ in Eq.\ (\ref{eq:calfdef}) ensures that the planar diagrams contain only contractions between $U$ (the $1,1$ block of ${\bf U}$) and $U^{\dagger}$ (the $2,2$ block of ${\bf U}$). In terms of the generating function $h$ we obtain the result
\begin{eqnarray} \label{eq:calfeq}
  {\bf F} &=& (2 z)^{-1} \sum_{\pm} \left( {\bf C} \pm {\bf C} {\bf T}' (1 - {\bf \Sigma}_{\pm} {\bf X}_{\pm})^{-1} {\bf \Sigma}_{\pm} {\bf T} {\bf C} z^{-1/2} \right), \\
  {\bf \Sigma}_{\pm} &=& \left( {\cal P} {\bf X}_{\pm} (1 - {\bf \Sigma}_{\pm} {\bf X}_{\pm})^{-1}\right)\, h \left( \left({\cal P} {\bf X}_{\pm} (1 - {\bf \Sigma}_{\pm} {\bf X}_{\pm})^{-1} \right)^{2} \right). \label{eq:calfeqB}
\end{eqnarray}

It remains to solve the $2 \times 2$ matrix equation (\ref{eq:calfeqB}). We could not do this analytically for arbitrary $\Gamma_j$, but only for the case of two identical tunnel barriers: $N_1 = N_2 = \case{1}{2} M \equiv N$, $\Gamma_{j} = \Gamma_{j+N}$ ($j=1,2,\ldots,N$). The solution of Eq.\ (\ref{eq:calfeqB}) in that case is
\begin{equation}
  {\bf \Sigma}_{\pm} = \pm \left(\sqrt{z} - \sqrt{z-1}\right) \left( \begin{array}{cc} 0 & \openone_M \\ \openone_M & 0 \end{array} \right),
\end{equation}
independent of the $\Gamma_j$'s. The trace of the Green function is
\begin{eqnarray}
  \mbox{tr}\, F(z) &=& \sum_{j=1}^{N} {2 (1 - \Gamma_j)(\sqrt{z} - \sqrt{z-1})  + \Gamma_j/\sqrt{z-1} \over 2 z (1 - \Gamma_j) (\sqrt{z} - \sqrt{z-1}) + \Gamma_j \sqrt{z}},
\end{eqnarray}
and the corresponding density of transmission eigenvalues is
\begin{eqnarray} \label{eq:rhogamma}
  \rho(T) = \sum_{j=1}^{N} {\Gamma_j(2-\Gamma_j) \over \pi (\Gamma_j^2 - 4 \Gamma_j T + 4 T)\sqrt{T(1-T)}}.
\end{eqnarray}
As a check, we note that $\rho(T) \to N \delta(T)$ if $\Gamma_j \to 0$ for all $j$, and $\rho(T) \to N \pi^{-1} [T(1-T)]^{-1/2}$ if $\Gamma_j \to 1$ for all $j$ [in agreement with Eq.\ (\ref{eq:rho})].

\section{Application to a Normal-metal--superconductor junction} \label{sec:NS}

\begin{figure}
\hspace{0.35\hsize}
\epsfxsize=0.3\hsize
\epsffile{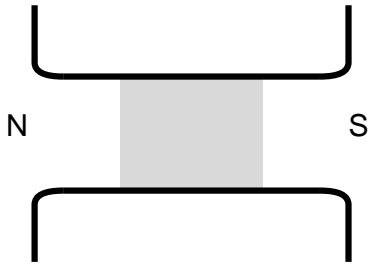}
\medskip

\caption{\label{fig:NS1} Conductor consisting of a normal metal (grey) coupled to one normal-metal reservoir (N) and one superconducting reservoir (S). The conductor may consist of a disordered segment or of a quantum dot.}
\end{figure}

As an altogether different application of the diagrammatic technique, we consider a junction between a normal metal (N) and a superconductor (S) (see Fig.\ \ref{fig:NS1}). At temperatures and voltages below the excitation gap $\Delta$ in S, conduction takes place via the mechanism of Andreev reflection \cite{Andreev}: An electron coming from N with an energy $\varepsilon$ (relative to the Fermi energy $E_F$) is reflected at the NS interface as a hole with energy $-\varepsilon$. The missing charge of $2e$ is absorbed by the superconducting condensate. We calculate the average and variance of the conductance, for the two cases that the NS junction consists of a disordered wire or of a chaotic cavity.

Starting point of the calculation is the relationship between the differential conductance $G_{\rm NS}(eV) = dI/dV$ of the NS junction and the transmission and reflection matrices of the normal region \cite{Beenakker92},
\begin{eqnarray}
  G_{\rm NS}(\varepsilon) &=& {4 e^2 \over h}\, \mbox{tr}\, \left( t'(\varepsilon) \left[1 + r'(-\varepsilon)^{*} r'(\varepsilon)\right]^{-1} t(-\varepsilon)^* \right)
        \left( t'(\varepsilon) \left[1 + r'(-\varepsilon)^{*} r'(\varepsilon)\right]^{-1} t(-\varepsilon)^{*} \right)^{\dagger}.
  \label{eq:NSconductance2}
\end{eqnarray}
This formula requires $eV \ll \Delta \ll E_F$ and zero temperature. The reflection and transmission matrices are $N \times N$ matrices, which together constitute the $2N \times 2N$ scattering matrix $S$. Using the polar decomposition (\ref{eq:polardecomp1}) we may rewrite the conductance formula (\ref{eq:NSconductance2}) as
\begin{eqnarray}
  G_{\rm NS}(\varepsilon) &=& 
    {4 e^2 \over h}
    \mbox{tr}\, \left[ T_{+}^{\vphantom{*}} 
      \left(1 + u_{+}^{\vphantom{*}} \sqrt{1-T_{-}^{\vphantom{*}}} u_{-}^*
        \sqrt{1-T_{+}^{\vphantom{*}}}\right)^{-1} u_{+}^{\vphantom{*}}
      \right. \nonumber \\ && \mbox{} \cdot \left. 
      T_{-}^{\vphantom{*}} u_{+}^{\vphantom{*}\dagger} 
      \left(1 + \sqrt{1-T_{+}^{\vphantom{*}}} u_{-}^{\rm T}
        \sqrt{1-T_{-}^{\vphantom{*}}} u_{+}^{\vphantom{*}\dagger}\right)^{-1} \right],
  \label{eq:NSconductance3}
\end{eqnarray}
where $T_{\pm} = T(\pm \varepsilon)$ and $u_{\pm} = w'(\pm \varepsilon) w(\mp \varepsilon)^{*}$. In the presence of spin-orbit scattering, $S$ is a matrix of quaternions, and the transpose should be replaced by the dual. In what follows, we will consider the case of no spin-orbit scattering. Spin-orbit scattering (considered by Slevin, Pichard, and Mello \cite{SlevinPichardMello}) will be included at the end by means of the translation rule of Sec.\ \ref{sec:trans}.

Averages are computed in two steps: first over the unitary matrix $u$, then over the matrix of transmission eigenvalues $T$. Four cases can be distinguished, depending on the magnitude of the magnetic field $B$ and voltage $V$ relative to the characteristic field $B_c$ for breaking time-reversal symmetry (${\cal T})$ and characteristic voltage $E_c/e$ for breaking electron-hole degeneracy (${\cal D}$) \cite{footnote4}:
\begin{itemize}
\item[0.] $eV \ll E_c$, $B \ll B_c$\ $\Longleftrightarrow$\ ${\cal T}$ and ${\cal D}$ are both present: Then $u_{\pm}$ may be approximated by the unit matrix, so that one only needs to average over the transmission eigenvalues. This case has been studied extensively \cite{BeenakkerLesHouches} and does not concern us here.
\item[1.] $eV \ll E_c$, $B \gg B_c$\ $\Longleftrightarrow$\ ${\cal D}$ is present, but ${\cal T}$ is broken: Then we may neglect the $\varepsilon$-dependence of $S$, so that $u_{+} = u_{-} \equiv u$. According to the isotropy assumption, $u$ is uniformly distributed in ${\cal U}(N)$.
\item[2.] $eV \gg E_c$, $B \ll B_c$\ $\Longleftrightarrow$\ ${\cal T}$ is present, but ${\cal D}$ is broken: Then we may consider $S(\varepsilon)$ and $S(-\varepsilon)$ as independent unitary symmetric matrices. Hence $u_{+}^{\vphantom{\dagger}} = u_{-}^{\dagger} \equiv u$ is uniformly distributed in ${\cal U}(N)$.
\item[3.] $eV \gg E_c$, $B \gg B_c$\ $\Longleftrightarrow$\ both ${\cal T}$ and ${\cal D}$ are broken: Then $u_{+}$ and $u_{-}$ are independent, both uniformly distributed in ${\cal U}(N)$.
\end{itemize}
We compute the average and variance of the conductance for cases 1, 2, and 3.

\subsection{Average conductance}

\begin{figure}
\hspace{0.15\hsize}
\epsfxsize=0.7\hsize
\epsffile{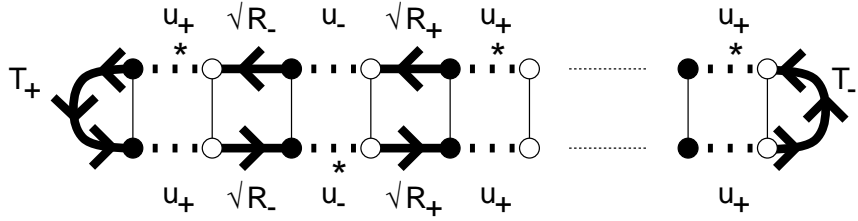}
\medskip

\caption{\label{fig:NSWL4} Ladder diagram for the ${\cal O}(N)$ contribution to $\langle G_{\rm NS} \rangle$. We defined $R_{\pm}=1-T_{\pm}$.}
\end{figure}

We start with the computation of the average conductance $\langle G_{\rm NS} \rangle$. We first perform the average $\langle \cdots \rangle_u$ over $u_{\pm}$ and then over $T_{\pm}$. To leading order only ladder diagrams contribute, see Fig.\ \ref{fig:NSWL4}. The result is the same for cases 1, 2, and 3:
\begin{mathletters} \label{eq:GNSavgu0}
\begin{eqnarray}
   \langle G_{\rm NS}/G_0 \rangle_{u} &=&
  2 N {{\tau}_{1+}^{\vphantom{2}} {\tau}_{1-}^{\vphantom{2}}
 \over {\tau}_{1+}^{\vphantom{2}} +
       {\tau}_{1-}^{\vphantom{2}} - 
       {\tau}_{1+}^{\vphantom{2}} {\tau}_{1-}^{\vphantom{2}}} +
       {\cal O}(1), \\
       {\tau}_{k\pm} &=& 
         {1 \over N}\, \mbox{tr}\, T_{\pm}^{k} = 
         {1 \over N} \sum_{j=1}^{N} T_j^{k}(\pm \varepsilon).
\end{eqnarray}
\end{mathletters}%
The ${\cal O}(1)$ contribution $\delta G_{\rm NS}$ is different for the three cases.

\begin{figure}
\hspace{0.15\hsize}
\epsfxsize=0.7\hsize
\epsffile{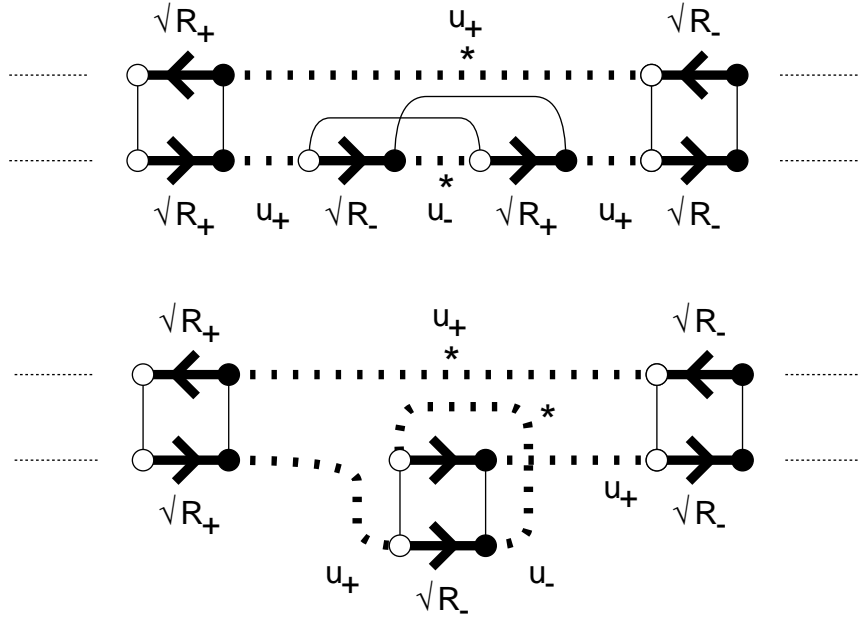}
\medskip

\caption{\label{fig:NSWL7} Maximally crossed diagram for the ${\cal O}(1)$ correction to $\langle G_{\rm NS} \rangle$ in the absence of time-reversal symmetry and presence of electron-hole degeneracy (top). The right and left parts of the diagram have a ladder structure. The central part may be redrawn as a ladder diagram (bottom).}
\end{figure}

\begin{figure}
\hspace{0.05\hsize}
\epsfxsize=0.9\hsize
\epsffile{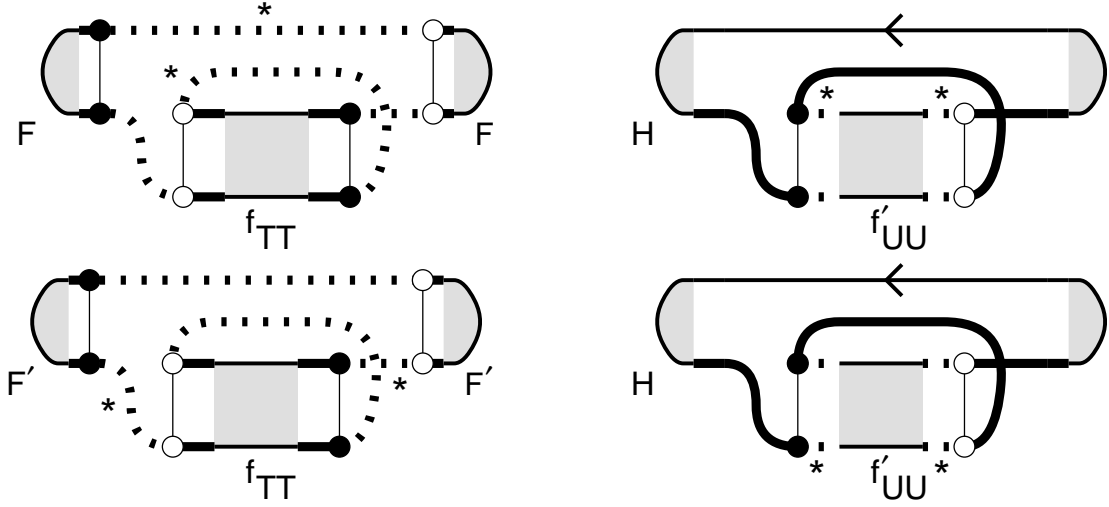}
\medskip

\caption{\label{fig:NSWL2} Diagrams for the ${\cal O}(1)$ correction to $\langle G_{\rm NS} \rangle$ in the absence of time-reversal symmetry and presence of electron-hole degeneracy.}
\end{figure}

Case 1, absence of ${\cal T}$ and presence of ${\cal D}$. We put $u_{\pm} = u$, $\tau_{k\pm} = \tau_k$. For normal metals, the ${\cal O}(1)$ contribution $\delta G$ to $\langle G \rangle$ vanishes if ${\cal T}$ is broken. However, in the NS junction an ${\cal O}(1)$ contribution remains \cite{BrouwerBeenakker95-1}. The diagrams which contribute to $\delta G_{\rm NS}$ have a maximally crossed central part, with contractions between $U$'s and $U^{*}$'s on the same side of the diagram (Fig.\ \ref{fig:NSWL7}, top). The left and right ends have a ladder structure. In the Hamiltonian approach, a similar maximally crossed diagram has been studied by Altland and Zirnbauer \cite{AltlandZirnbauer}, who call it a ``symplecton''. In total four diagrams contribute to $\delta G_{\rm NS}$, see Fig.\ \ref{fig:NSWL2}. The building blocks of the diagram have the algebraic expressions
\begin{mathletters} \label{eq:NSnotation}
\begin{eqnarray}
  F_{\pm} &=&
    T_{\pm} +
    (1-T_{\pm})\, \mbox{tr}\, T_{\pm}\,
      \mbox{tr}\, (1 - T_{\mp})\,
      \sum_{j=0}^{\infty} N^{-2j-2}
      \left[ \mbox{tr}\, (1-T_{+})\,
       \mbox{tr}\, (1 - T_{-})\right]^{j} \nonumber \\ &=&
    \left( {\tau_{1\pm} + T_{\pm} \tau_{1\mp} \ - 
      \tau_{1+} \tau_{1-}} \right)
    \left( {\tau_{1+} + \tau_{1-} -
      \tau_{1+} \tau_{1-}} \right)^{-1}, \\
  F'_{\pm} &=& 
    - (1 - T_{\mp})\, \mbox{tr}\, T_{\pm}\,
      \sum_{j=0}^{\infty} N^{-2j-1} 
      \left[ \mbox{tr}\, (1-T_{+})\,
        \mbox{tr}\, (1 - T_{-})\right]^{j} \nonumber \\ &=&
    - \left( {\tau_{1\pm} -
      \tau_{1\pm} T_{\mp}} \right)
   \left( { \tau_{1+} + \tau_{1-} -
        \tau_{1+} \tau_{1-}} \right)^{-1}, \\
  H_{\pm} &=&
    i N^{-1}  T_{\pm} \sqrt{1-T_{\pm}}\,
      \mbox{tr}\, F_{\mp} -
    i N^{-2} (1-T_{\pm})\sqrt{1-T_{\pm}}\, 
      \mbox{tr}\, F_{\mp}\, \mbox{tr}\, F'_{\pm}, \\
  f_{TT\pm} &=&
    - \mbox{tr}\, (1-T_{\pm})\,
      \sum_{j=0}^{\infty} N^{-2j} 
      \left[ \mbox{tr}\, (1-T_{+})\,
        \mbox{tr}\, (1 - T_{-})\right]^{j} \nonumber \\ &=&
    - N(1-\tau_{1\pm}) 
      \left({ \tau_{1+} + \tau_{1-} -
        \tau_{1+} \tau_{1-}} \right)^{-1}, \\
  f_{UU\pm} &=&
    - \mbox{tr}\, (1-T_{\pm})\,
     \sum_{j=0}^{\infty} N^{-2j-2}
     \left[ \mbox{tr}\, (1-T_{+})\,
       \mbox{tr}\, (1-T_{-})\right]^{j} \nonumber \\ &=&
    - N^{-1}(1 - \tau_{1\pm})
      \left[ {\tau_{1+} + \tau_{1-} -
        \tau_{1+} \tau_{1-}} \right]^{-1}, \\
  f_{UU\pm}' &=& 
    \sum_{j=0}^{\infty} N^{-2j-1}
      \left[ \mbox{tr}\, (1-T_{+})\,
      \mbox{tr}\, (1 - T_{-})\right]^{j} =
    N^{-1} \left( {\tau_{1+} + \tau_{1-} - 
        \tau_{1+} \tau_{1-}} \right)^{-1}.
\end{eqnarray}
\end{mathletters}%
Capital letters indicate matrices, lower-case letters indicate scalars. The subscripts $\pm$ are omitted from Fig.\ \ref{fig:NSWL2} because of electron-hole degeneracy. The ${\cal O}(1)$ correction $\delta G_{\rm NS}$ represented in Fig.\ \ref{fig:NSWL2} equals
\begin{eqnarray}
  \delta G_{\rm NS}/G_0 &=& 
     8 f_{UU}'\, \mbox{tr}\, i H \sqrt{1-T}\, + 4 W_{2} f_{TT} [(\mbox{tr}\, F)^2 + (\mbox{tr}\, F')^2]
     \nonumber \\ &=&
     - {8 \tau_1^{\vphantom{2}} - 4 \tau_1^2 + 4 \tau_1^3 - 8 \tau_2^{\vphantom{2}} \over 
        \tau_1 (2 - \tau_1)^3}. \label{eq:deltaGNSub2}
\end{eqnarray}

We still have to average over the transmission eigenvalues. We use that the sample-to-sample fluctuations $\tau_k - \langle \tau_k \rangle$ are an order $1/N$ smaller than the average. (This is a general property of a linear statistics, i.e.\ of quantities of the form $A = \sum_n a(T_n)$, see Ref.\ \onlinecite{BeenakkerReview}.) Hence
\begin{equation}
  \langle f({\tau}_k) \rangle = f(\langle {\tau}_k \rangle) [1 + {\cal
O}(N^{-2})], \label{eq:Nofluct}
\end{equation}
which implies that we may replace the average of the rational functions (\ref{eq:GNSavgu0}) and (\ref{eq:deltaGNSub2}) of the ${\tau}_k$'s by the rational functions of the average $\langle {\tau}_k \rangle$. This average has the $1/N$ expansion
\begin{equation}
  \langle {\tau}_k \rangle = \langle {\tau}_k \rangle_0 + {\cal O}(N^{-2}),
\label{eq:TauAvgExpand}
\end{equation}
where $\langle {\tau}_k \rangle_0$ is ${\cal O}(N^0)$. There is no term of order $N^{-1}$ in the absence of ${\cal T}$. The average over $T$ of Eqs.\ (\ref{eq:GNSavgu0}) and (\ref{eq:deltaGNSub2}) becomes
\begin{equation}
 \langle G_{\rm NS}/G_0 \rangle =
   {2N\langle {\tau}_1 \rangle_0^{\vphantom{2}} \over 2 -
\langle {\tau}_1 \rangle_0} - {8 \langle {\tau}_1 \rangle_0^{\vphantom{2}} - 4 \langle
{\tau}_1 \rangle_0^2 + 4\langle {\tau}_1 \rangle_0^3 - 8 \langle {\tau}_2
\rangle_0^{\vphantom{2}} \over \langle {\tau}_1 \rangle_0(2 - \langle {\tau}_1 \rangle_0)^3} +
{\cal O}(N^{-1}). \label{eq:GNSavgbeta2}
\end{equation}

\begin{figure}
\hspace{0.05\hsize}
\epsfxsize=0.9\hsize
\epsffile{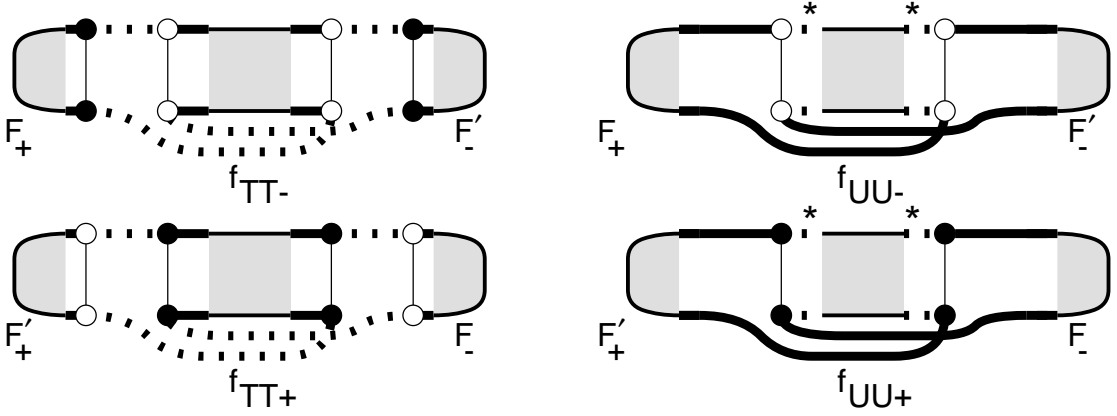}
\medskip

\caption{\label{fig:NSWL5} Diagrams for the ${\cal O}(1)$ correction to $\langle G_{\rm NS} \rangle $ in the absence of electron-hole degeneracy and presence of time-reversal symmetry.}
\end{figure}

Case 2, presence of ${\cal T}$ and absence of ${\cal D}$. We put $u_{+} = u_{-}^{\dagger} \equiv u$. The ${\cal O}(1)$ correction comes from the maximally crossed diagrams of Fig.\ \ref{fig:NSWL5},
\begin{eqnarray}
  \delta G_{\rm NS}/G_0 &=& 
    2 \, W_{2}\, \mbox{tr}\, F_{+}^{\vphantom{{\rm T}}}\, 
      f_{TT-}^{\vphantom{{\rm T}}}\, 
      \mbox{tr}\, F_{-}'^{\vphantom{{\rm T}}} +
    2 \, W_{2}\, \mbox{tr}\, F_{+}'^{\vphantom{{\rm T}}}\,
      f_{TT+}^{\vphantom{{\rm T}}}\,
      \mbox{tr}\, F_{-}^{\vphantom{{\rm T}}}
    \nonumber \\ && \mbox{} + 
    2 \, \mbox{tr}\, F_{+}^{\vphantom{{\rm T}}}
      f_{UU-}^{\vphantom{{\rm T}}} F_{-}'^{\rm T}\, +
    2 \, \mbox{tr}\, F_{+}'^{\vphantom{{\rm T}}}
      f_{UU+}^{\vphantom{{\rm T}}} F_{-}^{\rm T}.
\end{eqnarray}
Averaging over the transmission eigenvalues amounts to replacing $\tau_{k\pm}$ by its average, $\tau_{k\pm} \to \langle \tau_{k} \rangle_0 + N^{-1} \delta \tau_{k} + {\cal O}(N^{-2})$. (The average of $\tau_{k\pm}$ is the same for $+\varepsilon$ and $-\varepsilon$.) Because ${\cal T}$ is not broken there is a term of ${\cal O}(N^{-1})$ in this expression. We find for the average conductance
\begin{eqnarray}
  \langle G_{\rm NS}/G_0 \rangle &=& 
    {2 N \langle \tau_1 \rangle_0 \over 2 - \langle \tau_1 \rangle_0}
    \nonumber \\ && +
    {4\,  \delta {\tau}_1 \over (2 - \langle {\tau}_1 \rangle_0)^2} + 
    {4\,  \langle {\tau}_1 \rangle_0^2 - 4\, \langle {\tau}_1 \rangle_0^3 -
     4\,  \langle {\tau}_2 \rangle_0^{\vphantom{2}} +
     4\,  \langle {\tau}_1 \rangle_0^{\vphantom{2}}
       \langle {\tau}_2 \rangle_0^{\vphantom{2}} \over
       \langle {\tau}_1 \rangle_0\, (2 - \langle {\tau}_1 \rangle_0)^3}
  \label{eq:GNSavgbeta1V} + {\cal O}(N^{-1}).
\end{eqnarray}

Case 3, both ${\cal T}$ and ${\cal D}$ broken. Because $u_{+}$ and $u_{-}$ are independent, there are no diagrams which contribute to order $1$. The average conductance is obtained by averaging Eq.\ (\ref{eq:GNSavgu0}) over the transmission eigenvalues,
\begin{eqnarray} \label{eq:GNSavgbeta2V}
  \langle G_{\rm NS}/G_0 \rangle &=& 
    {2 N \langle \tau_1 \rangle_0 \over 2 - \langle \tau_1 \rangle_0} + {\cal O}(N^{-1}). 
\end{eqnarray}

From the translation rule of Sec.\ \ref{sec:trans} one deduces that in the presence of spin-orbit scattering, the leading ${\cal O}(N)$ term of the average conductance is unchanged, while the ${\cal O}(1)$ correction is multiplied by $-1/2$, in agreement with what was found by Slevin, Pichard and Mello \cite{SlevinPichardMello}.

The formulas given above apply to any system for which the isotropy assumption holds. We discuss two examples:

(a) A disordered wire (length $L$, mean free path $\ell$, number of transverse modes $N$), connected to a superconductor. We use the results \cite{MelloStone}
\begin{mathletters} \label{eq:DWprop}
\begin{eqnarray}
  \langle \tau_1 \rangle_0 &=& (1 + L/\ell)^{-1}, \\
  \langle \tau_2 \rangle_0 &=& 
    \case{2}{3} (1+L/\ell)^{-1} + \case{1}{3} (1+L/\ell)^{-4}, \\
  \delta \tau_1 &=& - \case{1}{3} (1+\ell/L)^{-3}.
\end{eqnarray}
\end{mathletters}%
We assume $\ell \ll L \ll N \ell$ and neglect terms of order $L/N\ell$ and $\ell/L$ but retain terms of order $1$ and $N \ell^p/L^p$ ($p \ge 1$). Substitution of Eq.\ (\ref{eq:DWprop}) into Eqs.\ (\ref{eq:GNSavgbeta2}), (\ref{eq:GNSavgbeta1V}), and (\ref{eq:GNSavgbeta2V}) yields
\begin{eqnarray} \label{eq:GavgWire}
  \langle G_{\rm NS}/G_0 \rangle =
  \left\{ \begin{array}{ll}
   N (1+ L/\ell)^{-1} - 1 + 4/\pi^2
                                 & (\mbox{${\cal D}$, ${\cal T}$}), \\
   N (1/2 + L/\ell)^{-1} - 1/3 & (\mbox{${\cal D}$, no ${\cal T}$}), \\
   N (1/2 + L/\ell)^{-1} - 2/3 & (\mbox{no ${\cal D}$, ${\cal T}$}), \\
   N (1/2 + L/\ell)^{-1}       & (\mbox{no ${\cal D}$, no ${\cal T}$}).
  \end{array} \right.
\end{eqnarray}
The result in the presence of both ${\cal T}$ and ${\cal D}$ has been taken from Refs.\ \onlinecite{Beenakker94,MacedoChalker}. In the presence of spin-orbit scattering, the ${\cal O}(N)$ term is unchanged, while the ${\cal O}(1)$ term is multiplied by $-1/2$.

(b) A chaotic cavity without tunnel barriers in the leads. Lead $1$ (with $N_1$ modes) is connected to a normal metal, lead $2$ (with $N_2$ modes) to a superconductor. An asymmetry between $N_1$ and $N_2$ appears because the dimension of $u_{\pm}$ in the polar decomposition (\ref{eq:polardecomp12}) is $N_2 \times N_2$. The $N_2 \times N_2$ matrix $T_{\pm}$ contains the $\mbox{min}\,(N_1,N_2)$ non-zero transmission eigenvalues on the diagonal (remaining diagonal elements being zero). We denote $N_{\rm tot} = N_1 + N_2$ and $N_{A} = ({N_1^2 + 6 N_1 N_2 + N_2^2})^{1/2}$. The averages $\langle \tau_1 \rangle_0$ and $\langle \tau_2 \rangle_0$ and the correction $\delta \tau_1$ can be computed from the density of transmission eigenvalues [Eqs.\ (\ref{eq:rho}) and (\ref{eq:drho})].  The results are
\begin{equation} \label{eq:momQD}
  \delta  \tau_{1}           = -N_1 N_2 N_{\rm tot}^{-2}, \ \
  \langle \tau_{1} \rangle_0 = N_1 N_{\rm tot}^{-1}, \ \
  \langle \tau_{2} \rangle_0 = N_1(N_{\rm tot}^{2} - N_1 N_2) N_{\rm tot}^{-3}.
\end{equation}
Substitution into Eqs.\ (\ref{eq:GNSavgbeta2}), (\ref{eq:GNSavgbeta1V}), and (\ref{eq:GNSavgbeta2V}) gives
\begin{eqnarray} \label{eq:GavgDot}
  \langle G_{\rm NS}/G_0 \rangle = \left\{ \begin{array}{ll}
  N_{\rm tot}^{\vphantom{2}} (1 - N_{\rm tot}^{\vphantom{2}}/N_{A}^{\vphantom{2}}) - 
    8 N_1^{\vphantom{2}} N_2^{\vphantom{2}} N_{\rm tot}^2/N_{A}^4 
    & (\mbox{${\cal D}$, ${\cal T}$}), \\
  2 N_1^{\vphantom{2}} N_2^{\vphantom{2}}/(N_{\rm tot}^{\vphantom{2}}+N_2^{\vphantom{2}}) - 4 N_1^{\vphantom{2}} N_2^{\vphantom{2}} N_{\rm tot}^{\vphantom{2}}/(N_{\rm tot}^{\vphantom{2}}+N_2^{\vphantom{2}})^3 
    & (\mbox{${\cal D}$, no ${\cal T}$}), \\
  2 N_1^{\vphantom{2}} N_2^{\vphantom{2}}/(N_{\rm tot}^{\vphantom{2}}+N_2^{\vphantom{2}}) - 4 N_2^{\vphantom{2}} N_{\rm tot}^2  /(N_{\rm tot^{\vphantom{2}}}+N_2^{\vphantom{2}})^3
    & (\mbox{no ${\cal D}$, ${\cal T}$}), \\
  2 N_1^{\vphantom{2}} N_2^{\vphantom{2}}/(N_{\rm tot}^{\vphantom{2}}+N_2^{\vphantom{2}})                        
    & (\mbox{no ${\cal D}$, no ${\cal T}$}).
  \end{array} \right.
\end{eqnarray}
The leading order term in Eq.\ (\ref{eq:GavgDot}) has also been obtained by Argaman and Zee \cite{Argaman}. (The case $N_1 = N_2$ was given in Ref.\ \onlinecite{JPB}).

\subsection{Conductance fluctuations} \label{sec:fluctNS}

To compute the variance of the conductance, we average in two steps: $\langle \cdots \rangle = \langle \langle \cdots \rangle_{{u}} \rangle_T$, where $\langle\cdots\rangle_{{u}}$ and $\langle\cdots\rangle_T$ are, respectively, the average over the unitary matrices ${{u_{\pm}}}$ and over the matrices of transmission eigenvalues ${T}_{\pm}$. It is convenient to add and subtract $\langle
\langle G_{\rm NS} \rangle^2_{{u}} \rangle_T^{\vphantom{2}}$, so that the
variance splits up into two parts,
\begin{eqnarray}
  \mbox{var}\, G_{\rm NS} &=&
    \left\langle \langle G_{\rm NS}^{\vphantom{2}}
      \rangle^2_{{u}} \right\rangle_T^{\vphantom{2}} -
    \left\langle \langle G_{\rm NS}^{\vphantom{2}} 
      \rangle_{{u}}^{\vphantom{2}} \right\rangle_T^2 +
    \left\langle \langle G_{\rm NS}^2 \rangle_{{u}}^{\vphantom{2}} -
      \langle G_{\rm NS}^{\vphantom{2}} \rangle^2_{{u}}
      \right\rangle_T^{\vphantom{2}},
\label{eq:VarGNSsplit}
\end{eqnarray}
which we evaluate separately.

The first two terms of Eqs.\ (\ref{eq:VarGNSsplit}) give the variance of $\langle G_{\rm NS}\rangle_{{u}}$ over the distribution of transmission eigenvalues. We calculated $\langle G_{\rm NS}\rangle_{{u}}$ in Eq.\ (\ref{eq:GNSavgu0}). Since $\langle G_{\rm NS} \rangle_{{u}}$ is a function of the linear statistic $\tau_{1\pm}$ only,  we know that its fluctuations are an order $1/N$ smaller than the average. This implies that, to leading order in $1/N$,
\begin{eqnarray}
  \left\langle \langle G_{\rm NS}^{\vphantom{2}}
      \rangle^2_{{u}} \right\rangle_T^{\vphantom{2}} -
    \left\langle \langle G_{\rm NS}^{\vphantom{2}} 
      \rangle_{{u}}^{\vphantom{2}} \right\rangle_T^2 &=&
     \sum_{\sigma,\sigma' = \pm}
      \left\langle {\partial \langle G_{\rm NS} \rangle_{u} \over
       \partial\, \tau_{1\sigma}} \right\rangle_T
      \left\langle {\partial \langle G_{\rm NS} \rangle_{u} \over 
       \partial\, \tau_{1\sigma'}} \right\rangle_T \,
     \mbox{covar}\,(\tau_{1\sigma}, \tau_{1\sigma'}) \nonumber \\ &=&
  8\, G_0^2\, N^2\, (2 - \langle \tau_1 \rangle_0)^{-4}\, \mbox{var}\, \tau_1\,
  \times\, \left\{ \begin{array}{ll}
    1 & (\mbox{without ${\cal D}$}), \\
    2 & (\mbox{with ${\cal D}$}). \end{array} \right.
  \label{eq:VarGNS1}
\end{eqnarray}

\begin{figure}
\hspace{0.1\hsize}
\epsfxsize=0.8\hsize
\epsffile{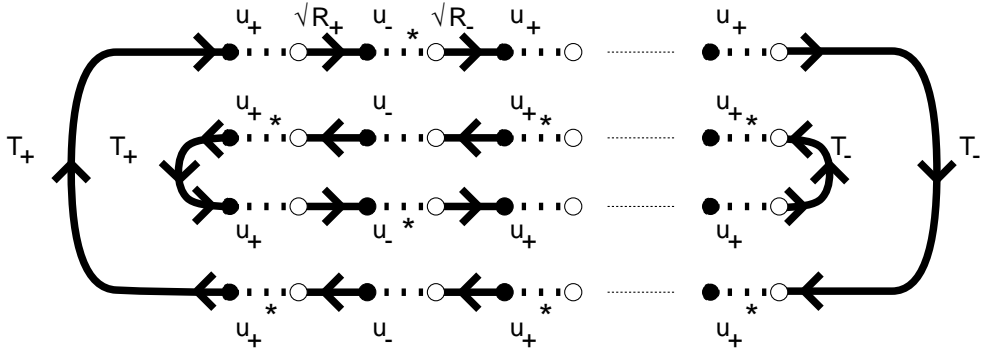}
\medskip

\caption{\label{fig:NSCF8} Diagrammatic representation of $G_{\rm NS}^2$.}
\end{figure}

We now turn to the third and fourth term of Eq.\ (\ref{eq:VarGNSsplit}). These terms involve the variance $\langle G_{\rm NS}^2 \rangle_{{u}}^{\vphantom{2}} - \langle G_{\rm NS}^{\vphantom{2}} \rangle^2_{{u}}$ of $G_{\rm NS}$ over ${\cal U}(N)$ and subsequently an average over the ${T}_n$'s. The calculation is similar to that of Sec.\ \ref{sec:fluctQD}. We represent $G_{\rm NS}^2$ by the diagram in Fig.\ \ref{fig:NSCF8}. The variance with respect to $u_{\pm}$ is given by the connected diagrams. We distinguish between two types of connected diagrams: (i) diagrams in which the inner and the outer loop are connected by a $T$-cycle or by a $U$-cycle, and (ii) diagrams in which the partition of the $U$-cycles involves a group which consists of a $U$-cycle from the inner loop and a $U$-cycle from the outer loop. The diagrams are similar to those of Fig.\ \ref{fig:QDCF7}, and are omitted. The final result is
\begin{eqnarray}
  \left\langle \langle G_{\rm NS}^2 \rangle_{{u}}^{\vphantom{2}} -
      \langle G_{\rm NS}^{\vphantom{2}} \rangle^2_{{u}} \right\rangle_T &=&
    8\, G_0^2\, \left(2 - \langle \tau_1 \rangle_0 \right)^{-6}
      \langle \tau_1 \rangle_0^{-2}
      \nonumber \\ && \mbox{} \times
      \left( 4  \langle \tau_1 \rangle_0^2 - 
        8 \langle \tau_1 \rangle_0^3 + 
        9 \langle \tau_1 \rangle_0^4 - 
        4 \langle \tau_1 \rangle_0^5 +
        2 \langle \tau_1 \rangle_0^6 -
        4 \langle \tau_1 \rangle_0  \langle \tau_2 \rangle_0
        \right. \nonumber \\ && \left. \mbox{}+
        2 \langle \tau_1 \rangle_0^2  \langle \tau_2 \rangle_0 - 
        2 \langle \tau_1 \rangle_0^3  \langle \tau_2 \rangle_0 -
        2 \langle \tau_1 \rangle_0^4  \langle \tau_2 \rangle_0 +
        6 \langle \tau_2 \rangle_0^2 - 
        6 \langle \tau_1 \rangle_0  \langle \tau_2 \rangle_0^2 
       \vphantom{\left[ \langle \rangle_{0}^{0} \right]} 
        \right. \nonumber \\ && \left. \mbox{} +
        \vphantom{\left[ \langle \rangle_{0}^{0} \right]}
        3 \langle \tau_1 \rangle_0^2 \langle \tau_2 \rangle_0^2 -
        4 \langle \tau_1 \rangle_0 \langle \tau_3 \rangle_0 +
        6 \langle \tau_1 \rangle^2 \langle  \tau_3 \rangle_0 -
        2 \langle \tau_1 \rangle_0^3 \langle \tau_3 \rangle_0 \right)
        \nonumber \\ && \mbox{} \times
        \left\{ \begin{array}{ll} 2 & (\mbox{${\cal D}$, no ${\cal T}$}), \\
                                  2 & (\mbox{${\cal T}$, no ${\cal D}$}), \\
                                  1 & (\mbox{no ${\cal D}$, no ${\cal T}$}).
        \end{array} \right. 
        \label{eq:VarGNS2}
\end{eqnarray}
The sum of Eqs.\ (\ref{eq:VarGNS1}) and (\ref{eq:VarGNS2}) equals $\mbox{var}\, G_{\rm NS}$, according to Eq.\ (\ref{eq:VarGNSsplit}).

In the presence of spin-orbit scattering $\mbox{var}\, G_{\rm NS}$ is four times as small, according to the translation rule of Sec.\ \ref{sec:trans}.

We give explicit results for the disordered wire and the chaotic cavity.

(a) For the disordered wire one has \cite{MelloStone,BeenakkerBuettiker} $\mbox{var}\, \tau_1 = \case{1}{15} N^{-2}$, $\langle \tau_k \rangle_0 = \case{1}{2} (\ell /L) \Gamma(\case{1}{2}) \Gamma(k) / \Gamma(k+\case{1}{2})$. Substitution into Eqs.\ (\ref{eq:VarGNS1}) and (\ref{eq:VarGNS2}) yields the variance
\begin{eqnarray}
  \mbox{var}\, G_{\rm NS}/G_0 =
  \left\{ \begin{array}{lll}
  16/15 - 48/\pi^4 & \approx 0.574 & (\mbox{${\cal D}$, ${\cal T}$}), \\
  8/15 & \approx 0.533 & (\mbox{${\cal D}$, no {${\cal T}$}}), \\
  8/15 & \approx 0.533 & (\mbox{${\cal T}$, no {${\cal D}$}}), \\
  4/15 & \approx 0.267 & (\mbox{no ${\cal D}$, no {${\cal T}$}}).
  \end{array} \right.
\end{eqnarray}
The result in the presence of both ${\cal T}$ and ${\cal D}$ has been taken from Ref.\ \onlinecite{MacedoChalker,BeenakkerRejaei}. If both ${\cal D}$ and ${\cal T}$ are present, breaking {${\cal T}$} (or ${\cal D}$) reduces the variance by less than $10$\% \cite{BrouwerBeenakker95-2,Marmorkos}.

(b) For the chaotic cavity one has $\mbox{var}\, \tau_1 = 2 N_1^2/\beta N_{\rm tot}^4$ and $\langle \tau_{3} \rangle_0 = N_1^{\vphantom{2}} (N_{\rm tot}^4 - 2 N_{\rm tot}^2 N_1^{\vphantom{2}} N_2^{\vphantom{2}} + 2 N_1^2 N_2^2)/N_{\rm tot}^5$ [see Eqs.\ (\ref{eq:QDvarC}) and (\ref{eq:rho})]. In combination with Eq.\ (\ref{eq:momQD}) this gives
\begin{eqnarray}
  \mbox{var}\, G_{\rm NS}/G_0 =
  \left\{ \begin{array}{ll}
  128 N_1^2 N_2^2 (N_{\rm tot}^4 + 2 N_1^2 N_2^2)(N_{\rm tot}^2 + 4 N_1^{\vphantom{2}} N_2^{\vphantom{2}})^{-4} & 
    (\mbox{${\cal D}$, ${\cal T}$}), \\
  32 N_2^2 N_{\rm tot}^2 (N_{\rm tot}^2 - N_1^{\vphantom{2}} N_2^{\vphantom{2}}) (N_{\rm tot}^{\vphantom{2}} + N_2^{\vphantom{2}})^{-6} & 
   (\mbox{${\cal D}$, no {${\cal T}$}}), \\
 32 N_2^2 N_{\rm tot}^2 (N_{\rm tot}^2 - N_1^{\vphantom{2}} N_2^{\vphantom{2}}) (N_{\rm tot}^{\vphantom{2}} + N_2^{\vphantom{2}})^{-6} & 
   (\mbox{${\cal T}$, no {${\cal D}$}}), \\
 16 N_2^2 N_{\rm tot}^2 (N_{\rm tot}^2 - N_1^{\vphantom{2}} N_2^{\vphantom{2}}) (N_{\rm tot}^{\vphantom{2}} + N_2^{\vphantom{2}})^{-6} & 
   (\mbox{no ${\cal D}$, no {${\cal T}$}}).
  \end{array} \right.
 \label{eq:VarNSDot}
\end{eqnarray}
If the coupling between the cavity and the normal metal is weak compared to the coupling to the superconductor ($N_2 \gg N_1$), one finds $\mbox{var}\, G_{\rm NS}(\mbox{${\cal D}$, ${\cal T}$})/\mbox{var}\, G_{\rm NS}(\mbox{${\cal D}$, no ${\cal T}$}) =  {\cal O}(N_1/N_2)^2$. In this case breaking ${\cal T}$ greatly enhances the conductance fluctuations. In the opposite case, if the couplings are equal ($N_1 = N_2$), one finds $\mbox{var}\, G_{\rm NS}(\mbox{${\cal D}$, ${\cal T}$})/\mbox{var}\, G_{\rm NS}(\mbox{${\cal D}$, no ${\cal T}$}) = 2187/2084 \approx 1.07$. In this case breaking ${\cal T}$ has almost no effect on the conductance fluctuations.

\section{Summary} \label{sec:SUMM}

We developed a diagrammatic technique for the evaluation of integrals of polynomial functions of unitary matrices over the unitary group ${\cal U}(N)$. In the large-$N$ limit the number of relevant diagrams is restricted, which allows for the evaluation of integrals over rational functions. We also considered integrals of unitary symmetric matrices, by means of a slight modification of the diagrammatic rules. A translation rule was given to relate integrals of (self-dual) unitary matrices of quaternions to integrals over (symmetric) unitary matrices of complex numbers.

We discussed two applications: A chaotic cavity (quantum dot) with tunnel barriers in the leads and a normal-metal--superconductor (NS) junction. In both cases, the conductance is a rational function of a unitary matrix. In the large-$N$ limit the average conductance is given by a series of ladder diagrams. The weak-localization correction consists of maximally-crossed diagrams. These two types of diagrams are analogous to the diffuson and cooperon diagrams in the diagrammatic perturbation theory for disordered systems \cite{Anderson,Gorkov}. We computed the density of transmission eigenvalues, where the leading order term is given by planar diagrams. Resummation of the diagrams leads to a Dyson equation for the Green function, similar to that encountered in the theory of integrals over Hermitian matrices \cite{Pandey,BrezinZee}.

For the NS junction, the ${\cal O}(1)$ correction to the average conductance is non-zero in the presence of a magnetic field, because of a different type of maximally crossed diagrams. These diagrams are suppressed by a sufficiently large voltage to break electron-hole degeneracy. The new type of maximally crossed diagrams explains the coexistence of weak localization with a magnetic field \cite{BrouwerBeenakker95-1} and the insensitivity of the conductance fluctuations to a magnetic field \cite{BrouwerBeenakker95-2,Marmorkos}.

This research was supported by the ``Ne\-der\-land\-se or\-ga\-ni\-sa\-tie voor
We\-ten\-schap\-pe\-lijk On\-der\-zoek'' (NWO) and by the ``Stich\-ting voor
Fun\-da\-men\-teel On\-der\-zoek der Ma\-te\-rie'' (FOM).

\appendix

\section{Weight factors for polynomial integrals} \label{app:coeff}

In Tables \ref{tab:2} -- \ref{tab:5} we list the weight factors $V_{c_1,\ldots,c_k}$ and $W_{c_1,\ldots,c_k}$ for $n = c_1 + \ldots + c_k \le 5$ for the CUE and the COE. (Tables of $V$ are also given in Refs.\ \onlinecite{Samuel,Mello} for the CUE and in Ref.\ \onlinecite{MelloSeligman} for the COE.) The weight factors are rational functions of the dimension $N$ of the unitary matrix. The denominators $A_{n}$ and $B_{n}$ of, respectively, $V_{c_1,\ldots,c_k}$ and $W_{c_1,\ldots,c_k}$ depend only on $n$. They are tabulated in Tables \ref{tab:2} and \ref{tab:3}. The numerators $A_n V_{c_1,\ldots,c_k}$ and $B_n W_{c_1,\ldots,c_k}$ are tabulated in Tables \ref{tab:4} and \ref{tab:5}.

\begin{table}
\begin{center}
\begin{tabular}{c|c|c}
  $n$     & $A_{n}$ (CUE) & $A_{n}$ (COE) \\ \hline
  $1$ & $N$ & $N+1$ \\
  $2$ & $N(N^2-1)$ & $N(N+1)(N+3)$ \\
  $3$ & $N(N^2-1)(N^2-4)$ & $(N-1)N(N+1)(N+3)(N+5)$ \\
  $4$ & $N^2(N^2-1)(N^2-4)(N^2-9)$ & $(N-2)(N-1)N(N+1)(N+2)(N+3)$ \\ && $\mbox{} \times (N+5)(N+7)$ \\
  $5$ & $N^2(N^2-1)(N^2-4)(N^2-9)(N^2-16)$ & $(N-3)(N-2)(N-1)N(N+1)(N+2)$ \\ && $\mbox{} \times (N+3)(N+5)(N+7)(N+9)$
\end{tabular}
\end{center}

\caption{\label{tab:2} Denominators $A_n$ of the coefficients $V_{c_1,\ldots,c_k}$ for $n = c_1 + \ldots + c_k \le 5$.}
\end{table}

\begin{table}
\begin{center}
\begin{tabular}{c|c|c}
 $c_1,\ldots,c_k$ & $A_n V_{c_1,\ldots,c_k}$ (CUE) & $A_n V_{c_1,\ldots,c_k}$ (COE) \\ \hline
 $1$ &         $1$ & $1$ \\ \hline
 $1,1$ &       $N$ & $2+N$ \\
 $2$ &         $-1$ & $-1$ \\ \hline
 $1,1,1$ &     $-2 + {N^2}$ & $2 + 5\,N + {N^2}$ \\
 $2,1$ &       $-N$ & $-3 - N$ \\
 $3$ &         $2$ & $2$ \\ \hline
 $1,1,1,1$ &   $6 - 8\,{N^2} + {N^4}$ & $-32 - 8\,N + 28\,{N^2} + 11\,{N^3} + {N^4}$ \\
 $2,1,1$ &     $4\,N - {N^3}$ & $-4 - 18\,N - 9\,{N^2} - {N^3}$ \\
 $2,2$ &       $6 + {N^2}$ & $24 + 7\,N + {N^2}$ \\
 $3,1$ &       $-3 + 2\,{N^2}$ & $10 + 12\,N + 2\,{N^2}$ \\
 $4$ &         $-5\,N$ & $-11 - 5\,N$ \\ \hline
 $1,1,1,1,1$ & $78\,N - 20\,{N^3} + {N^5}$ & $128 - 408\,N - 84\,{N^2} + 59\,{N^3} + 16\,{N^4} + {N^5}$ \\
 $2,1,1,1$ &   $-24 + 14\,{N^2} - {N^4}$ & $92 + 38\,N - 43\,{N^2} - 14\,{N^3} - {N^4}$ \\
 $2,2,1$ &     $-2\,N + {N^3}$ & $56 + 43\,N + 12\,{N^2} + {N^3}$ \\
 $3,1,1$ &     $-18\,N + 2\,{N^3}$ & $-52 + 40\,N + 22\,{N^2} + 2\,{N^3}$ \\
 $3,2$ &       $-24 - 2\,{N^2}$ & $-88 - 18\,N - 2\,{N^2}$ \\
 $4,1$ &       $24 - 5\,{N^2}$ & $-7 - 36\,N - 5\,{N^2}$ \\
 $5$ &         $14\,N$ & $38 + 14\,N$ \\
\end{tabular}
\end{center}

\caption{\label{tab:4} Numerators $A_n V_{c_1,\ldots,c_k}$ of the coefficients $V_{c_1,\ldots,c_k}$ for $n = c_1 + \ldots + c_k \le 5$. The denominators $A_n$ are given Table \protect\ref{tab:2}.}
\end{table}

\begin{table}
\begin{center}
\begin{tabular}{c|c|c}
  $n$     & $B_{n}$ (CUE) & $B_{n}$ (COE) \\ \hline
  $1$ & $N$ & $N+1$ \\
  $2$ & $N^2(N^2-1)$ & $N(N+1)^2(N+3)$ \\
  $3$ & $N^3(N^2-1)(N^2-4)$ & $(N-1)N(N+1)^3(N+3)(N+5)$ \\
  $4$ & $N^4(N^2-1)^2(N^2-4)(N^2-9)$ & $(N-2)(N-1)N^2(N+1)^4(N+2)$ \\ && $\mbox{} \times (N+3)^2(N+5)(N+7)$ \\
  $5$ & $N^5(N^2-1)^2(N^2-4)(N^2-9)(N^2-16)$ & $(N-3)(N-2)(N-1)N^2(N+1)^5(N+2)$ \\ && $\mbox{} \times (N+3)^2(N+5)(N+7)(N+9)$
\end{tabular}
\end{center}

\caption{\label{tab:3} Denominators $B_n$ of the coefficients $W_{c_1,\ldots,c_k}$ for $n = c_1 + \ldots + c_k \le 5$.}
\end{table}
\begin{table}
\begin{center}
\begin{tabular}{c|c|c}
 $c_1,\ldots,c_k$ & $B_{n} W_{c_1,\ldots,c_k}$ (CUE) & $B_{n} W_{c_1,\ldots,c_k}$ (COE) \\ \hline
 $1$ &          $1$ & $1$ \\ \hline
 $1,1$ &        $1$ & $2$ \\
 $2$ &          $-N$ & $-1-N$ \\ \hline
 $1,1,1$ &      $8$ & $32$ \\
 $2,1$ &        $-4\,N$ & $-8 - 8\,N$ \\
 $3$ &          $2\, N^2$ & $2 + 4\,N + 2\,{N^2}$ \\ \hline
 $1,1,1,1$ &    $-216 + 144\,{N^2}$ & $-1680 + 6720\,N + 6096\,{N^2} + 1152\,{N^3}$ \\
 $2,1,1$ &      $72\,N - 48\,{N^3}$ & $280 - 840\,N - 2136\,{N^2} - 1208\,{N^3} - 192\,{N^4}$ \\
 $2,2$ &        $-42\,{N^2} + 18\,{N^4}$ & $-140 - 116\,N + 384\,{N^2} + 592\,{N^3} + 268\,{N^4} + 36\,{N^5}$ \\
 $3,1$ &        $-15\,{N^2} + 15\,{N^4}$ & $198\,N + 552\,{N^2} + 540\,{N^3} + 216\,{N^4} + 30\,{N^5}$ \\
 $4$ &          $5\,{N^3} - 5\,{N^5}$ & $-33\,N - 125\,{N^2} - 182\,{N^3} - 126\,{N^4} - 41\,{N^5} - 5\,{N^6}$ \\ \hline
 $1,1,1,1,1$ &  $-13824 + 4224\,{N^2}$ & $-483840 + 297984\,N + 407040\,{N^2} + 67584\,{N^3}$ \\
 $2,1,1,1$ &    $3456\,N - 1056\,{N^3}$ & $60480 + 23232\,N - 88128\,{N^2} - 59328\,{N^3} - 8448\,{N^4}$ \\
 $2,2,1$ &      $-1248\,{N^2} + 288\,{N^4}$ & $-12096 - 21120\,N + 1152\,{N^2} + 18432\,{N^3} + 9408\,{N^4} + 1152\,{N^5}$ \\
 $3,1,1$ &      $-480\,{N^2} + 240\,{N^4}$ & $-3024 + 192\,N + 15072\,{N^2} + 18432\,{N^3} + 7536\,{N^4} + 960\,{N^5}$ \\
 $3,2$ &        $312\,{N^3} - 72\,{N^5}$ & $1512 + 4152\,N + 2496\,{N^2} - 2448\,{N^3} - 3480\,{N^4} - 1320\,{N^5} - 144\,{N^6}$\\
 $4,1$ &        $56\,{N^3} - 56\,{N^5}$ & $-912\,N - 3376\,{N^2} - 4768\,{N^3} - 3168\,{N^4} - 976\,{N^5} - 112\,{N^6}$ \\
 $5$ &          $-14\,{N^4} + 14\,{N^6}$ & $114\,N + 536\,{N^2} + 1018\,{N^3} + 992\,{N^4} + 518\,{N^5} + 136\,{N^6} + 14\,{N^7}$\\
\end{tabular}
\end{center}

\caption{\label{tab:5} Numerators $B_n W_{c_1,\ldots,c_k}$ of the coefficients $W_{c_1,\ldots,c_k}$ for $n = c_1 + \ldots + c_k \le 5$. The denominators $B_n$ are given in Table \protect\ref{tab:3}.}
\end{table}


\begin{references}
%
%
\bibitem{WeidenmuellerReview} H. A. Weidenm\"uller, Physica A {\bf 167}, 28 (1990).
\bibitem{StoneReview} A. D. Stone, P. A. Mello, K. A. Muttalib, and J.-L. Pichard, in {\em Mesoscopic Phenomena in Solids}, edited by B. L. Al'tshuler, P. A. Lee, and R. A. Webb (North--Holland, Amsterdam, 1991).
\bibitem{MelloReview} P. A. Mello, in {\em Mesoscopic Quantum Physics}, edited by E. Akkermans, G. Montambaux, J.-L. Pichard, and J. Zinn-Justin (North-Holland, Amsterdam, 1995).
\bibitem{BeenakkerReview} C. W. J. Beenakker, Rev.\ Mod.\ Phys., to be published.
\bibitem{BarangerMello} H. U. Baranger and P. A. Mello, Phys.\ Rev.\ Lett.\ {\bf 73}, 142 (1994).
\bibitem{JPB} R. A. Jalabert, J.-L. Pichard, and C. W. J. Beenakker, Europhys.\ Lett.\ {\bf 27}, 255 (1994).
\bibitem{Dyson} F. J. Dyson, J.\ Math.\ Phys.\ {\bf 3}, 140 (1962); {\bf 3}, 157 (1962).
\bibitem{BlumelSmilansky} R. Bl\"umel and U. Smilansky, Phys.\ Rev.\ Lett.\
{\bf 60}, 477 (1988); {\bf 64}, 241 (1990); U. Smilansky, in {\em
Chaos and Quantum Physics}, edited by M.-J. Giannoni, A. Voros, and J.
Zinn-Justin (North-Holland, Amsterdam, 1991).
\bibitem{Dorokhov} O. N. Dorokhov, Pis'ma Zh. Eksp. Teor. Fiz. {\bf 36}, 259 (1982) [JETP Lett. {\bf 36}, 318 (1982)].
\bibitem{MPK} P. A. Mello, P. Pereyra, and N. Kumar, Ann. Phys. (N.Y.) {\bf 181}, 290 (1988).
\bibitem{Pastur} L. A. Pastur, Teoret.\ Mat.\ Fiz.\ {\bf 10}, 102 (1972) [Theoret.\ Math.\ Phys.\ {\bf 10}, 67 (1972)].
\bibitem{Pandey} A. Pandey, Ann.\ Phys.\ (N.Y.) {\bf 134}, 110 (1981).
\bibitem{Efetov} K. B. Efetov, Adv.\ Phys.\ {\bf 32}, 53 (1983).
\bibitem{VWZ} J. J. M. Verbaarschot, H. A. Weidenm\"uller, and M. R. Zirnbauer, Phys. Rep. {\bf 129}, 367 (1985).
\bibitem{IWZ} S. Iida, H. A. Weidenm\"uller, and J. A. Zuk, Phys.\ Rev.\ Lett.\ {\bf 64}, 583 (1990); Ann. Phys. (N.Y.) {\bf 200}, 219 (1990).
\bibitem{PrigodinEfetovIida} V. N. Prigodin, K. B. Efetov and S. Iida, Phys. Rev.\ Lett.\ {\bf 71}, 1230 (1993); Phys.\ Rev.\ B {\bf 51}, 17223 (1995). 
\bibitem{BrezinZee} E. Br\'ezin and A. Zee, Phys.\ Rev.\ E {\bf 49}, 2588 (1994).
\bibitem{MMZ} A. D. Mirlin, A. M\"uller-Groeling, and M. R. Zirnbauer, Ann. Phys. (N.Y.) {\bf 236}, 325 (1994).
\bibitem{BrouwerBeenakker94} P. W. Brouwer and C. W. J. Beenakker, Phys.\ Rev.\ B {\bf 50}, 11263 (1994); {\bf 51}, 7739 (1995).
\bibitem{BarangerMello96} H. U. Baranger and P. A. Mello, Phys.\ Rev.\ B {\bf 51}, 4703 (1995); Europhys.\ Lett.\ {\bf 33}, 465 (1996).
\bibitem{FriedmanMello2} W. A. Friedman and P. A. Mello, Ann. Phys. (N.Y.) {\bf 161}, 276 (1985).
\bibitem{Anderson} P. W. Anderson, E. Abrahams, and T. V. Ramakrishnan, Phys.\ Rev.\ Lett.\ {\bf 43}, 718 (1979).
\bibitem{Gorkov} L. P. Gor'kov, A. I. Larkin, and D. E. Khmel'nitski\v{\i}, Pis'ma Zh.\ Eksp.\ Teor.\ Fiz.\ {\bf 30}, 248 (1979) [JETP Lett.\ {\bf 30}, 228
1979].
\bibitem{Altshuler} B. L. Al'tshuler, Pis'ma Zh.\ Eksp.\ Teor.\ Fiz.\ {\bf 41}, 530 (1985) [JETP Lett.\ {\bf 41}, 648 (1985)].
\bibitem{LeeStone} P. A. Lee and A. D. Stone, Phys.\ Rev.\ Lett.\ {\bf 55}, 1622 (1985); P. A. Lee, A. D. Stone, and H. Fukuyama, Phys.\ Rev.\ B {\bf 35}, 1039 (1987).
\bibitem{BrouwerBeenakker95-1} P. W. Brouwer and C. W. J. Beenakker, Phys.\ Rev.\ B {\bf 52}, R3868 (1995). 
\bibitem{AltlandZirnbauer} A. Altland and M. R. Zirnbauer, preprints (cond-mat/9508026, cond-mat/9602137).
\bibitem{BrouwerBeenakker95-2} P. W. Brouwer and C. W. J. Beenakker, Phys.\ Rev.\ B {\bf 52}, 16772 (1995).
\bibitem{Argaman} N. Argaman and A. Zee, preprint (cond-mat/9603136).
%
%
\bibitem{Creutz} M.\ Creutz, J.\ Math.\ Phys.\ {\bf 19}, 2043 (1978).
\bibitem{Samuel} S. Samuel, J.\ Math.\ Phys.\ {\bf 21}, 2695 (1980).
\bibitem{Mello} P. A. Mello, J.\ Phys.\ A {\bf 23}, 4061 (1990).
\bibitem{Mehta} M. L. Mehta, {\em Random Matrices} (Academic, New York, 1991).
%
%
\bibitem{MelloSeligman} P. A. Mello and T. H. Seligman, Nucl.\ Phys.\ A {\bf 344}, 489 (1980).
%
%
\bibitem{Juengling} K. J\"ungling and R. Oppermann, Z.\ Phys.\ B {\bf 38}, 93 (1980); R. Oppermann and K. J\"ungling, Phys.\ Lett.\ A {\bf 76}, 449 (1980).
\bibitem{Wegner} F. Wegner, Z.\ Phys.\ B {\bf 94}, 297 (1983).
%
%
\bibitem{Hua}  L.\ K.\ Hua, {\em Harmonic Analysis of Functions of
Several Complex Variables in the Classical Domains} (Amer.\ Math.\ Soc.,\
Providence, 1963).
\bibitem{MPS} P. A. Mello, P. Pereyra, and T. H. Seligman,
Ann.\ Phys.\ (N.Y.) {\bf 161}, 254 (1985). 
\bibitem{DoronSmilansky} E. Doron and U. Smilansky, Nucl.\ Phys.\ A {\bf 545}, 455 (1992).
\bibitem{Brouwer} P. W. Brouwer, Phys.\ Rev.\ B {\bf 51}, 16878 (1995).
\bibitem{NazarovR} Yu. V. Nazarov, in {\em Quantum Dynamics of Submicron Structures}, edited by H. A. Cerdeira, B. Kramer, and G. Sch\"{o}n, NATO ASI Series E291 (Kluwer, Dordrecht, 1995).
\bibitem{Nazarov} Yu. V. Nazarov, Phys. Rev. Lett. {\bf 73}, 134 (1994).
%
%
\bibitem{Andreev} A. F. Andreev, Zh.\ Eksp.\ Teor.\ Fiz.\ {\bf 46}, 1823 (1964) [Sov.\ Phys.\ JETP {\bf 19}, 1228 (1964)].
\bibitem{Beenakker92} C. W. J. Beenakker, Phys.\ Rev.\ B {\bf 46}, 12841 (1992).
\bibitem{SlevinPichardMello} K. Slevin, J.-L. Pichard, and P. A. Mello, J.\ Phys.\ I (France) {\bf 6}, 529 (1996).
\bibitem{footnote4} In a disordered wire (length $L$, width $W$, mean free path $\ell$), one has $B_c = h/eLW$, $E_c = \hbar v_F \ell/L^2$. In a chaotic cavity (area $A$, mean dwell time $\tau$, mean time to cross the cavity $\tau'$) one has $B_c = (h/eA) (\tau'/\tau)^{1/2}$, $E_c = \hbar/\tau$).
\bibitem{BeenakkerLesHouches} For a review, see: C. W. J. Beenakker, in  {\em Mesoscopic Quantum Physics}, edited by E. Akkermans, G. Montambaux, J.-L. Pichard, and J. Zinn-Justin (North-Holland, Amsterdam, 1995).
\bibitem{MelloStone} P. A. Mello and A. D. Stone, Phys.\ Rev.\ B {\bf 44}, 3559 (1991).
\bibitem{Beenakker94} C. W. J. Beenakker, Phys.\ Rev.\ B {\bf 49}, 2205
(1994). 
\bibitem{MacedoChalker} A. M. S. Mac\^edo and J. T. Chalker, Phys.\ Rev.\ B
{\bf 49}, 4695 (1994). 
\bibitem{BeenakkerBuettiker} C. W. J. Beenakker and M. B\"uttiker, Phys.\ Rev.\ B {\bf 46}, 1889 (1992).
\bibitem{BeenakkerRejaei} C. W. J. Beenakker and B. Rejaei, Phys.\ Rev.\
B {\bf 49}, 7499 (1994).
\bibitem{Marmorkos} I. K.  Marmorkos, C. W. J.  Beenakker, and R. A.  Jalabert, Phys.\ Rev.\ B {\bf 48}, 2811 (1993).
\end{references}
\end{document}